\documentclass[aps,twocolumn,amsmath,amssymb,footinbib,showpacs,longbibliography,superscriptaddress]{revtex4-1}

\usepackage{times}

\usepackage[english]{babel}
\usepackage{latexsym}
\usepackage{graphics}
\usepackage{subfigure}
\usepackage{epsfig}
\usepackage{color}
\usepackage{soul}
\usepackage{hyperref}
\usepackage{braket} 
\usepackage[T1]{fontenc}
\usepackage[latin9]{inputenc}
\usepackage{amstext}
\usepackage{amssymb}
\usepackage{graphicx}
\usepackage{esint}
\usepackage{braket}
\usepackage{babel}
\usepackage{amsmath}

\usepackage{mathrsfs}
\usepackage{ulem}

\hypersetup{
colorlinks=true,
citecolor=blue,
linkcolor=red,
urlcolor=black
}


\newcommand{\ie}{i.e.}

\begin{document}

\title{
Probing universal phase diagram of dimensional crossover with an atomic quantum simulator
}

\author{Jinyuan Tian}
\homepage{These authors contributed equally to this work.}
\author{Zhongcheng Yu}
\homepage{These authors contributed equally to this work.}
\affiliation{State Key Laboratory of Photonics and Communications, School of Electronics, Peking University, Beijing 100871, China}
\author{Jing Liu}
\affiliation{Institute of Advanced Functional Materials and Devices, Shanxi University, Taiyuan 030031, China}
\affiliation{Institute of Carbon-based Thin Film Electronics, Peking University, Shanxi, Taiyuan 030012, China}
\author{Chi-Kin Lai}
\affiliation{State Key Laboratory of Photonics and Communications, School of Electronics, Peking University, Beijing 100871, China}
\author{Lorenzo Pizzino}
\affiliation{DQMP, University of Geneva, 24 Quai Ernest-Ansermet, CH-1211 Geneva, Switzerland}
\author{Chengyang Wu}
\affiliation{State Key Laboratory of Photonics and Communications, School of Electronics, Peking University, Beijing 100871, China}
\author{Hongmian Shui}
\affiliation{State Key Laboratory of Photonics and Communications, School of Electronics, Peking University, Beijing 100871, China}
\affiliation{Institute of Carbon-based Thin Film Electronics, Peking University, Shanxi, Taiyuan 030012, China}
\author{Thierry Giamarchi}
\email{thierry.giamarchi@unige.ch}
\affiliation{DQMP, University of Geneva, 24 Quai Ernest-Ansermet, CH-1211 Geneva, Switzerland}
\author{Hepeng Yao}
\email{hepeng.yao@pku.edu.cn}
\affiliation{State Key Laboratory of Photonics and Communications, School of Electronics, Peking University, Beijing 100871, China}
\author{Xiaoji Zhou}
\email{xjzhou@pku.edu.cn}
\affiliation{State Key Laboratory of Photonics and Communications, School of Electronics, Peking University, Beijing 100871, China}
\affiliation{Institute of Carbon-based Thin Film Electronics, Peking University, Shanxi, Taiyuan 030012, China}

\date{\today}

%%%%%%%%%%%%%%%%%%%%%%%%%%%%%%%%%
\begin{abstract}
%Dimensionality is a crucial parameter in scientific studies. In quantum physics, reducing dimensionality usually enhances the interaction and fluctuations, giving rise to novel properties. Thanks to these, quantum simulators at low dimensions have been widely used in recent years. However, a full phase diagram of dimensionality and temperature is still lacking. Here, we generate a quantum simulator with high tunability of dimensionality and temperature, and measure its full phase diagram. We locate quantum regimes from three to zero dimensions, and find the thermal phase emerges surprisingly between them. For different dimensionalities, we also study the quantum-to-thermal transition which follows different properties. Especially, we find a special regime where a quantum dimensional crossover happens by just increasing temperature. Our results shed light on further research about low-dimensional and coupled condensed matter structures which are crucial for studies about organic and high temperature superconductivity.

Dimensionality is a fundamental concept in physics, which plays a hidden but crucial role in various domains, including condensed matter physics, relativity and string theory, statistical physics, etc. In quantum physics, reducing dimensionality usually enhances fluctuations and leads to novel properties. 
Owing to these effects, quantum simulators in which dimensionality can be controlled have emerged as a new area of interest. However, such a platform has only been studied in specific regimes and a universal phase diagram is lacking. 
Here, we produce an interacting atomic quantum simulator with continuous tunability of anisotropy and temperature, and probe the universal phase diagram of dimensional crossover. At low temperatures, we identify the regimes from quantum three to zero dimensions. By increasing temperature, we observe the non-trivial emergence of a thermal regime situated between the quantum zero and integer dimensions. We show that the quantum-to-thermal transition falls into four different universality classes depending on the dimensionality.
Surprisingly, we also detect a fifth type where the high-dimensional quantum system can reach the thermal phase by crossing a low-dimensional quantum regime. Our results provide a crucial foundation for understanding the projective condensed matter structures in unconventional dimensions.

\end{abstract}
%%%%%%%%%%%%%%%%%%%%%%%%%%%%%%%%%

\maketitle

%%%%%%%%%%%%%%%%%%%%%%%%%%%%%%%%%
%%%%%%%%%%%%%%%%%%%%%%%%%%%%%%%%%
%\section{Introduction}

%Dimensionality is a fundamental concept which plays a crucial role in various research domains, including chemical reaction, DNA replication, crustal evolution, phase transition and etc. 

The Euclidian space we live in is three-dimensional (3D), where the equations of states or motions are well defined and studied. In recent years, research objects with dimensionality different from three are widely found naturally or artificially. They usually lead to novel physical properties, quite different from the ones of the 3D world. For instance, relativity and string theory, which plays an important role in particle physics and cosmology, mostly rely on dimensions larger than three~\cite{wald2010general}. In condensed matter physics, fractal structures can play an important role. They are usually characterized by the so-called fractal dimension, which is non-integer~\cite{akkermans2013,jagannathan2021fibonacci}.
%Statistical physics suggests certain rules donnot apply to low dimensions.

In microscopic physics, be it classical or quantum, the role of dimensionality is also essential.
Even for a classical system which can be described by the Boltzmann distribution, the probability density function of kinetic energy exhibits totally different dependence between three and low dimensions~\cite{PATHRIA2022}. In quantum physics, the distinction is even stronger owing to the very different properties of quantum fluctuation in different dimensions. 
Various types of high-temperature and organic superconductors show novel properties arising from reduced dimensionality~\cite{doi:10.1126/science.288.5465.468,doi:10.1021/cr030647c,CRPHYS_2024__25_G1_17_0}. However, controlling parameters like the tunneling rates along different directions (e.g. with pressure or chemistry) accurately and continuously, remains challenging~\cite{10.1093/oso/9780198562696.001.0001,PhysRevB.71.075104}.
Therefore, the atomic quantum simulators, mainly realized by ultracold atoms in optical potentials, has been widely extended to low dimensional structures in recent years owing to high controllability~\cite{bloch-review-2008,hadzibabic-2Dgas-2011,cazalilla-1dreview-2011}. 
In one or two dimensions, they reveal remarkable phenomena, such as the fermionization of bosons~\cite{paredes_tonks_experiment,kinoshita2004}, Tomonaga-Luttinger Liquid (TLL)~\cite{giamarchi_book_1d} type of correlation, topological properties~\cite{goldman-topology-2013,tarnowski-topology-2019}, frustrated phase~\cite{doi:10.1126/science.1207239} and the Berezinskii-Kosterlitz-Thouless (BKT) transition~\cite{hadzibabic2006,ha-strong2Dboson-2013}.
Although most systems are firmly based in one of the integer dimensions, some systems can, as a function of parameters, show behavior pertaining to several dimensionalities, known as dimensional crossover. In recent experiments, this phenomenon was analyzed using quantum simulators based on atomic~\cite{dalibard-crossoverD-2023,spielman-crossover-2023,Guo2024} and photonic systems~\cite{umesh-crossoverD-light-2024} in certain regimes. In these simulators, special behaviors of superfluidity and quantum correlation are observed, which reflect properties of multiple dimensions. 
%These simulators can be beneficial for the study of their projective structures in condensed matter physics, such as organic and high-temperature superconductors \cite{Guo2020,doi:10.1126/science.288.5465.468,doi:10.1021/cr030647c,Uji2001,book_Lebed2008,Jerome2004,Nomoto2023}, thanks to their high tunability of anisotropy. 

%The special properties of these quantum simulators can be beneficial for producing novel quantum devices, as well as the study of their projective structures in condensed matter physics, such as organic and high-temperature superconductors. 

%The special properties of these quantum simulators can be beneficial for the study of their projective structures in condensed matter physics. 

%The special properties of these quantum simulators can be beneficial for producing novel quantum devices and materials, as well as the study of their projective structures in condensed matter physics. 

\begin{figure*}
    \centering
    \includegraphics[width=2\columnwidth]{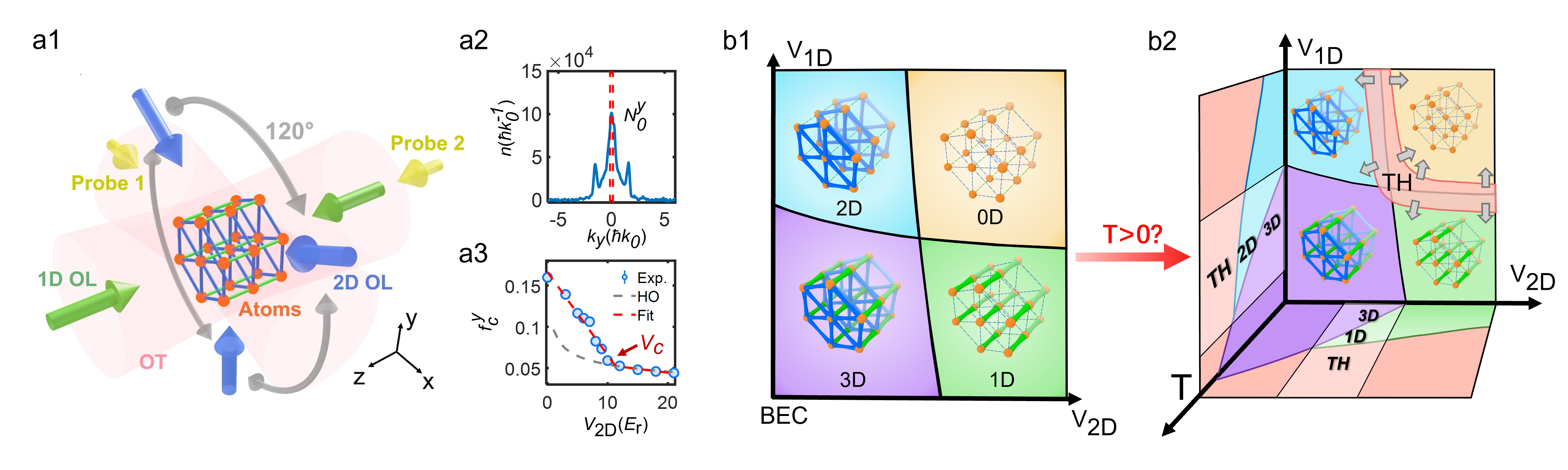}
    \caption{\textbf{Illustration of the experiment.}
    (a1) Sketch of the BEC system loading into a laser potential consisting of crossover optical dipole trap (OT, red cylinders), 1D optical lattice (1D OL, green arrows, z direction) and 2D triangular optical lattices (OL, blue arrows, x-y plane). The gravity is along the y-direction, while 1D lattice is aligned along the z-direction.
    The yellow arrows show the two probes.
    (a2) The momentum distribution from the TOF images for the case with lattice depths $V_{\rm 2D}=5.0{E_{\rm r}}$ and $V_{\rm 1D}=0.0{E_{\rm r}}$, and temperature $T=23 ~{\rm nK}$. The red dashed lines mark the zero-momentum area, which contains $N_0^y$ atoms. 
    (a3) The typical behavior of $f^y_c$ (blue circles) as a function of $V_{\rm 2D}$ with the same $T$ and $V_{\rm 1D}$ as (a2). The red dashed line is the piecewise fit which decides the critical potential $V_c$. The grey dashed line is $f^y_c$ computed by harmonic trap approximation.
    (b1) The sketch for the phase diagram of dimensional crossover at zero temperature,
    where the quantum 3D (purple), 2D (blue), 1D (green) and 0D (yellow) regimes are presented.
    In each phase, the subplot depicts the structural diagram, where lattice sites (orange sphere) are connected by coupling (blue and green lines), be it coherent(solid) or incoherent(dashed).
    (b2) The sketch for the finite-temperature behavior observed in this work. For fixed temperature, we find a thermal phase (TH) appears between zero and positive dimensions. When increasing temperature for fixed anisotropy, we find four common quantum-to-thermal transitions and one special type, where the system reaches the thermal phase via low-dimensional quantum regimes, such as 3D-1D-TH. 
     } 
    \label{fig1}
\end{figure*}

Temperature and interaction are two important parameters for the dimensional crossover of the quantum simulators. At zero temperature, the mechanism is clear~\cite{ho04_deconfined_bec,bloch-review-2008}. For a quantum system at dimension $D$, providing a constraint on $D'$ directions will suppress the tunneling rate along them and produce a system with dimensionality $D-D'$, see the sketch in Figs.~\ref{fig1}(a1) and (b1). Moving to finite temperature with zero interactions, temperature will simply provoke dimensional crossover when it coincides with the kinetic energy in the transverse directions $D'$. However, the mechanism with both finite temperature and interactions is non-trivial, since the tunneling between the blocks is not of the free-particle type any more. Previous works have only carried out studies for such a system in specific regimes or for specific quantities, theoretically~\cite{giamarchi_book_1d,Cazalilla_2006,PhysRevB.102.195145,yao-crossoverD-2022,Pizzino2024} and experimentally~\cite{dalibard-crossoverD-2023,spielman-crossover-2023,Guo2024,Jin_2019}. However, a study which reflects the universal properties of dimensional crossover for such a quantum simulator is lacking. Especially, the universal nature of the finite-temperature phase diagram for an interacting system is not clear.

In this work, we provide the first probe of the universal phase diagram of dimensional crossover with an atomic quantum simulator. Loading ultracold atomic system into triangular optical lattices, we obtain an interacting simulator with high tunability of anisotropy and temperature. At tens of nano-Kelvin, we can identify quantum regimes at different dimensionalities. By measuring the detailed phase diagram at different temperatures, two important universal features appear, see Fig.~\ref{fig1}(b2). On the one hand, for each fixed temperature, the thermal (TH, classical) regime always appears between 0D and positive integer-D quantum regimes, which can be explained by the interplay of quantum and thermal fluctuations. On the other hand, by increasing temperature for fixed anisotropy, we find the quantum-to-thermal transitions falling into different universality classes for different dimensionalities, namely the BEC transition (3D), BKT transition (2D), TLL transition (1D) and melting effect of Mott insulator (0D) accordingly. Strikingly, we also detect a new type of transition different from these four. For some special cases, the quantum 3D system can reach the thermal state via a low dimensional quantum phase, instead of a direct transition. This suggests that, by increasing temperature, a dimensional crossover between quantum systems may happen before the thermal transition. Our experimental data are in good agreement with quantum Monte Carlo.

%\section{Sketch of the core physics and experimental setup}

Our experiment starts from a Rb-87 Bose-Einstein Condensate (BEC) in the hyperfine state $F=1$ trapped in a crossed optical dipole trap containing typically $2.5 \times 10^5$ atoms~\cite{PhysRevLett.126.035301}, see Fig.~\ref{fig1}(a1). Its 3D s-wave scattering length is $a_{3D} = 107(4) a_0$. %, the BEC is in the Thomas-Fermi regime. 
By properly adjusting the magneto-optical trap (MOT) loading time before evaporative cooling, we can control the system's temperature from 16 nK up to 455 nK without significantly altering the atomic number. 
Then, we adiabatically ramp up a 3D optical lattice with lattice spacing $a=\lambda /2=532 ~{\rm nm}$ in $80$ ms and hold it for $20$ ms. As shown in Fig.~\ref{fig1}(a1), our optical lattices consist of a 2D triangular lattice ($x$-$y$ plane, blue arrows) parallel to the direction of gravity ($y$ direction) and a 1D lattice ($z$ direction, green arrows) perpendicular to it. The laser beams for the 1D and 2D lattices have a frequency difference of 110 MHz to ensure no interference between their laser beams. 
The lattice depths of the 2D triangular lattice $V_{\rm 2D}$ and the 1D constrained lattice $V_{\rm 1D}$ range from 0 to 25 $E_{\rm r}$ and 0 to 70 $E_{\rm r}$, respectively, with an accuracy of 0.2\%, where $E_{\rm r} = \pi^2 \hbar^2 / (2m a^2)$ is the recoil energy, with $\hbar$ the reduced Planck's constant and $m$ the mass of particles.

\begin{figure*}
    \centering
    \includegraphics[width=2\columnwidth]{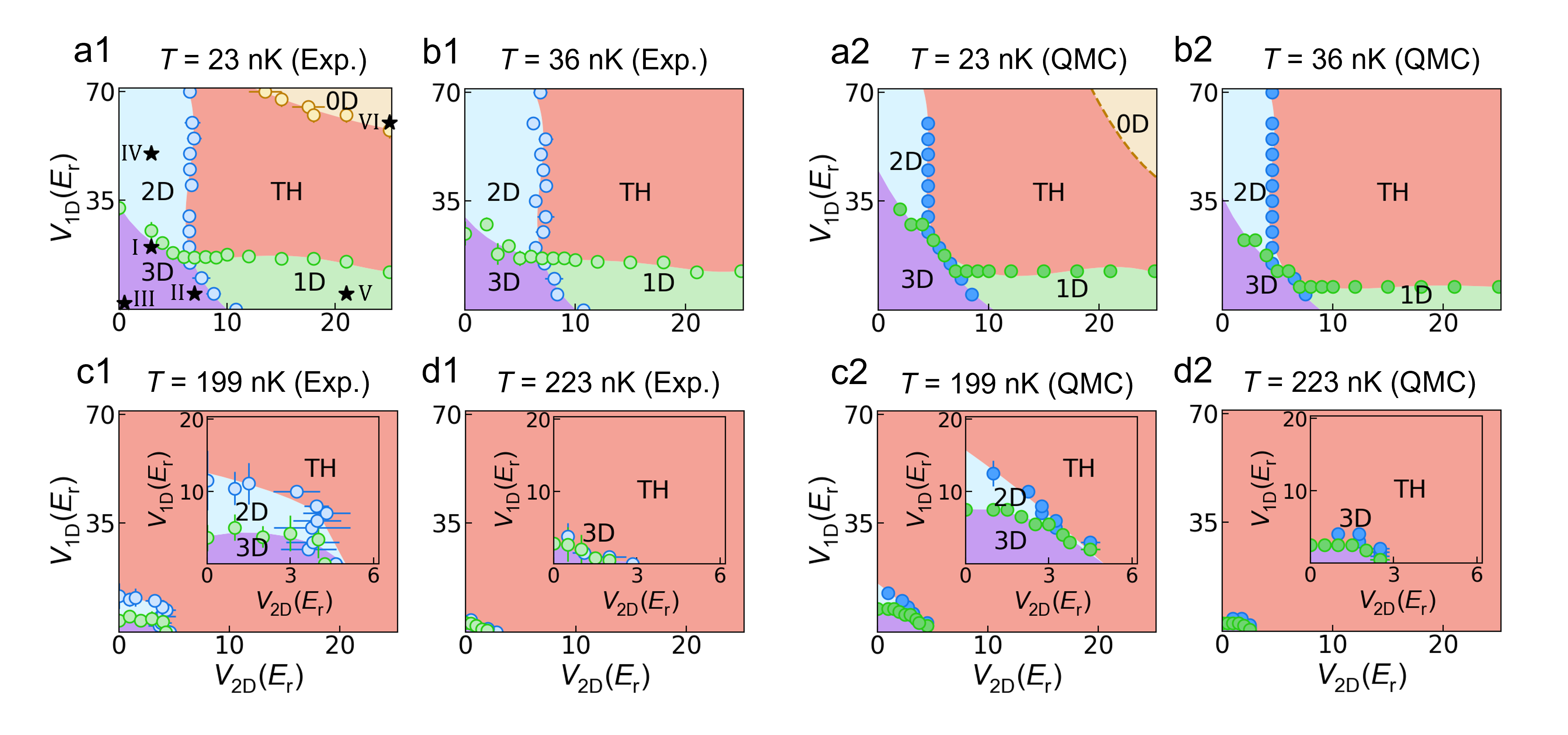}
    \caption{ \textbf{The universal phase diagram of dimensional crossover at different temperatures.} 
 (a1)-(a4) are the experimentally measured phase diagrams at initial BEC temperatures $T= 23(5)$, $36(3)$, $199(25)$ and $223(29)$ nK, as a function of the lattice amplitudes $V_{\rm 2D}$ and $V_{\rm 1D}$. 
Here, we observe quantum regimes at 3D (purple), 2D (blue), 1D (green), 0D (yellow) as well as the thermal regime (TH, red).
 The transition points are judged by zero-momentum fraction (blue and green circles) and correlation function (yellow circles). Insets in (c1) and (d1) are a zoom around the low-lattice-depth area. Error bars are obtained from the piecewise fit as in Fig.~\ref{fig1} (a3). The experimental parameters are particle number $N = 2.0(3) \times 10 ^5$ and 3D s-wave scattering length $a_{3D} = 107(4) a_0$ with trap frequencies  $(\omega_x,\omega_y,\omega_z)/2\pi$ ranging from (27,84,80) Hz to (60,135,121) Hz depending on the temperature considered. (a2)-(d2) are the counterpart for (a1)-(d1), which presents the QMC simulations for equivalent homogeneous systems. 
 }
\label{fig2} 
\end{figure*}

The experimental sequence starts from preparing the BEC at the targeted temperature. Then, we independently tune the depths of the 2D triangular and 1D lattices to tune the anisotropy of the system.  After holding it for another $20$ ms, we make sure the system reaches an equilibrium state~\cite{Greiner2002}. Next, we remove both the optical dipole traps and lattices for $28$ ms time-of-flight (TOF) and take the absorption image. With the images from the two probes along the $x$ and $z$ directions (yellow arrows in Fig.~\ref{fig1}(a1)), we observe the momentum distribution from which we extract useful information. More specifically, as suggested by Refs.~\cite{PhysRevA.84.061606,Guo2024,yao-crossoverD-2022,Pizzino2024}, we access the zero-momentum fraction along a certain direction. For instance, for the $y$ direction, it is defined as
\begin{equation}
    f^y_c = \frac{\int_{-\infty}^{+\infty}dk_x\int_{-\infty}^{+\infty}dk_z \int_{-\Delta k_y}^{+\Delta k_y} n(k) dk_y}{\int_{-\infty}^{+\infty}dk_x\int_{-\infty}^{+\infty}dk_z \int_{-\infty}^{+\infty} n(k) dk_y},
\end{equation}
with $n(k)$ the momentum distribution, $\Delta k_y=2\pi/L_y$ the zero-momentum width and $L_y$ the system size. Such a quantity reflects the quantum coherence properties along certain directions. It is more accurate than the visibility of the momentum diffraction peak~\cite{Kato2008} and has been proved to be efficient for studying the dimensional crossover~\cite{Guo2024,yao-crossoverD-2022,Pizzino2024}.   
From the typical momentum distribution as in Fig.~\ref{fig1}(a2), we compute the $f_c^y$ and construct Fig.~\ref{fig1}(a3). Clearly, we can apply a piecewise fit (dashed red line) and determine the critical lattice depth $V_c$ (see details in Supplementary Information). For systems at low temperatures, we always find that the large enough potential regime fits nicely with that of a harmonic oscillator (dashed grey line). %This agrees with the observation in Ref.~\cite{Guo2024} and proves the validity of our procedure to extrapolate the values of $V_c$. 

At zero temperature, the physics of such a system is qualitatively clear, see Fig.~\ref{fig1}(b1). When both $V_{\rm 2D}$ and $V_{\rm 1D}$ are small, the system is a 3D BEC with modulated density. When increasing $V_{\rm 1D}$ ($V_{\rm 2D}$ resp.) while keeping $V_{\rm 2D}$ ($V_{\rm 1D}$ resp.) small, the coupling along the $z$ direction ($x,y$ directions resp.) becomes incoherent, while it remains coherent along the others.
This is the regime of the 2D (1D resp.) atomic quantum simulator.
%This regime has attracted a lot of interest as platform to simulate 2D quantum systems, see for instance Refs.~\cite{hadzibabic2006,ha-strong2Dboson-2013,goldman-topology-2013,tarnowski-topology-2019}. Similarly, when increasing $V_{\rm 1D}$ while keeping $V_{\rm 2D}$ small, the system turns into 1D quantum tubes and many studies have been carried also in this direction~\cite{paredes_tonks_experiment,kinoshita2004,billy2008,derrico2014,yao-disorder-2024,vijayan-spin-charge-2020,Hulet-spin-2022}. 
In the limit where both $V_{\rm 2D}$ and $V_{\rm 1D}$ are large, the system becomes effectively 0D. All sites are decoupled due to quantum fluctuations and this is equivalent to the 3D Mott-insulator regime observed in Refs.~\cite{Greiner2002}.
However, as mentioned above, the interacting systems at finite temperatures remain unclear and form the central focus.

%%%%%%%%%%%%%%%%%%%%%%%%%%%%%%%%%
%\section{The phase diagram of dimensionality}

We first prepare our interacting BEC at different values of initial finite temperature $T$ and measure the phase diagram as a function of $V_{\rm 2D}$ and $V_{\rm 1D}$, see Fig.~\ref{fig2} for four typical cases. The blue and green circles are the critical lattice depths along the two directions respectively, judged by $f^y_c$ and $f^z_c$.  Notably, we benefit from the use of a triangular lattice which allows us to more easily tune the effective dimension of the system (see details in Supplementary materials).
Here we always load the lattice adiabatically and thus each diagrams is isentropic. At the lowest temperature we realized, \ie, $23$ nK, we find the four quantum regimes at different dimensionalities as predicted in Fig.~\ref{fig1}(b1). 
To further locate the thermal regime, we scan the zero-momentum fraction and correlation length as a function of $T$ for a large scale of data points, and check when these quantities saturate at a small value (yellow circles, see details below).
Interestingly, we find the thermal regime appears to be located between the zero- and positive integer dimensional quantum regimes. 
This behavior can be explained by the different effects of thermal fluctuations.
For the 0D system, it is an incompressible insulator with finite gaps whose correlation length is independent of temperature. The quantum-to-thermal transition for such a system is the melting of the gap~\cite{PhysRevA.71.063601,PhysRevLett.125.060401}, leading to a compressible thermal phase with a $T$-dependent correlation length. Thus, systems with smaller gap, \ie, smaller lattice amplitude,  will be melted first.
On the other hand, for quantum systems at positive integer dimensionalities, the quantum-to-thermal transition is induced by the joint contribution of quantum and thermal fluctuations. When the effective dimensionality is larger, \ie, smaller lattice amplitude, the quantum fluctuation is smaller and it calls for a higher temperature to enter the thermal phase~\cite{bloch-review-2008,cazalilla-1dreview-2011,hadzibabic-2Dgas-2011,pitaevskii_becbook}.
Thanks to the two processes mentioned above, the thermal phase appears in the middle of the phase diagram as in Fig.~\ref{fig2}(a1). As the temperature increases, the quantum regimes shrink successively and the thermal regime expands, see Fig.~\ref{fig2}(a1)-(d1). 
We find that the 0D, 1D, 2D, 3D quantum regimes disappear at the temperature of $T=36$ nK (b1), $199$ nK (c1), $223$ nK (d1) and $250$ nK, correspondingly. This fits with our previous statement.

To further confirm our observations, we run QMC simulations of an equivalent homogeneous system (see details in Method) and generate the phase diagram by studying the superfluid stiffness, see Fig.~\ref{fig2}(a2)-(d2). We qualitatively recover the same experimental phase diagrams. %with less than 25$\%$ difference. 
The quantitative discrepancy might be due to different factors such as the presence of the harmonic trap and the variation of the number of particles.

%%%%%%%%%%%%%%%%%%%%%%%%%%%%%%%%%
%\section{The dimensional crossover below quantum degeneracy}

\begin{figure}[t!]
    \includegraphics[width=1 \columnwidth]{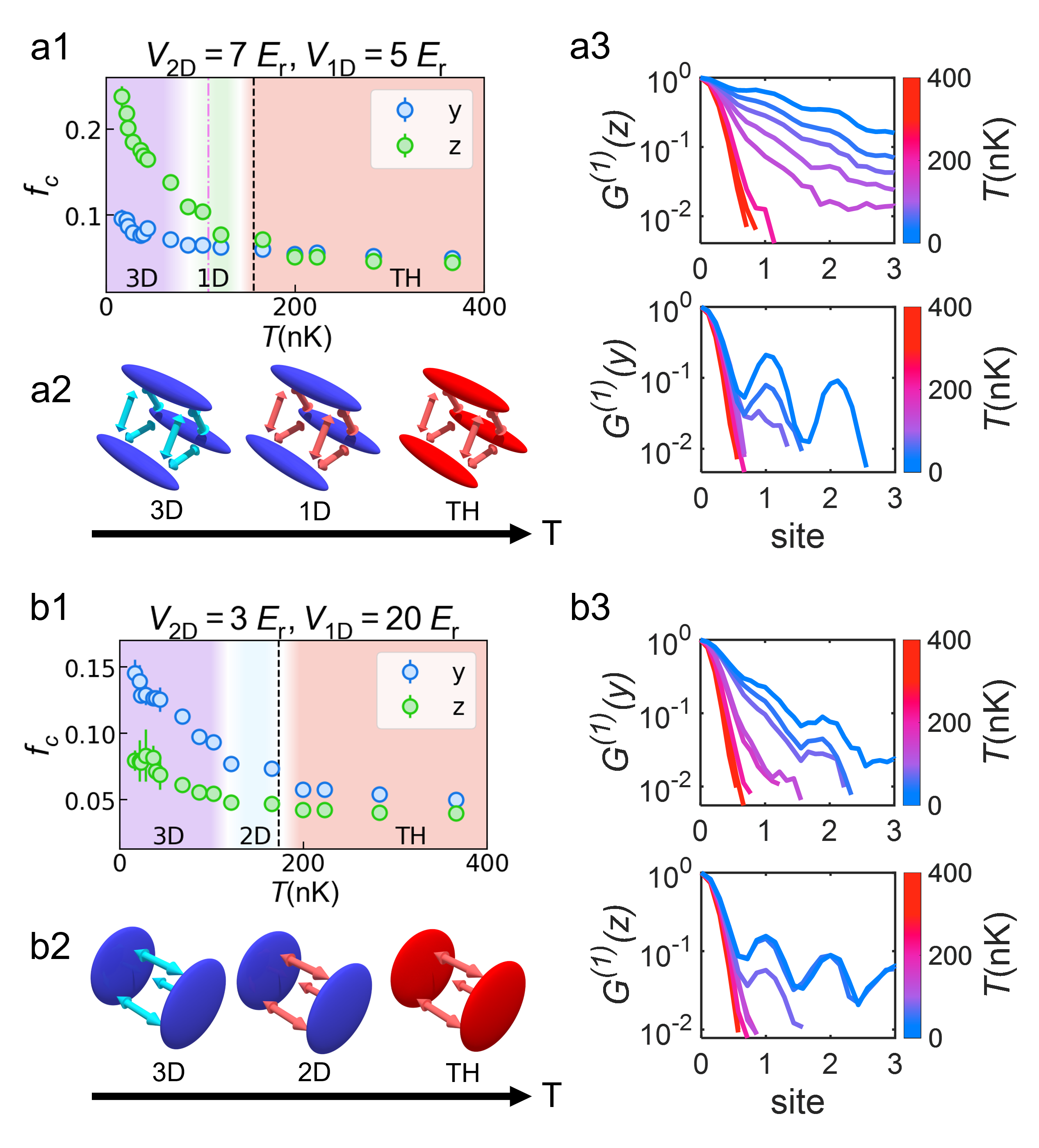}
    \caption{ \textbf{Special category of finite temperature transition} We show the zero-momentum fraction along two directions $f^y_c$ (blue circles) and $f^z_c$ (green circles) as a function of temperature $T$, for two different cases:
    (a1) $V_{\rm 2D}=3.0 ~{E_{\rm r}}$, $V_{\rm 1D}=20.0 ~{E_{\rm r}}$ and (b1) $V_{\rm 2D}=7.0 ~{E_{\rm r}}$, $V_{\rm 1D}=5.0 ~{E_{\rm r}}$.
    Error bars represent the standard deviation of five measurements. The colored areas represent the judged regimes. The white region is the estimated transition temperature from piecewise fit of experimental data and its with represent the errorbars.
    (a2) and (b2) are illustration for the physical pictures, where blue (red resp.) layers or tubes indicate quantum (thermal resp.) gases. Light blue (red resp.) legs indicate coherent (incoherent resp.) coupling. (a3) and (b3) are the corresponding correlation function $G^{(1)}$ along $y$ and $z$ directions at different temperatures. Experimental parameters: particle number $N = 2.0(3) \times 10 ^5$ and 3D s-wave scattering length $a_{3D} = 107(4) a_0$. }
    %For 3D state, there is a long-range correlation in both the $y$ and $z$ directions. For thermal states, the correlation decay along both directions drops exponentialy fast. While for 2D or 1D case, one can only observe coherence in the $y$ or $z$ direction, and the $G^{(1)}$ function in the other direction is the same as the thermal state.}
    \label{fig3}
\end{figure}

\begin{figure}
    \centering
    \includegraphics[width=1.03\columnwidth]{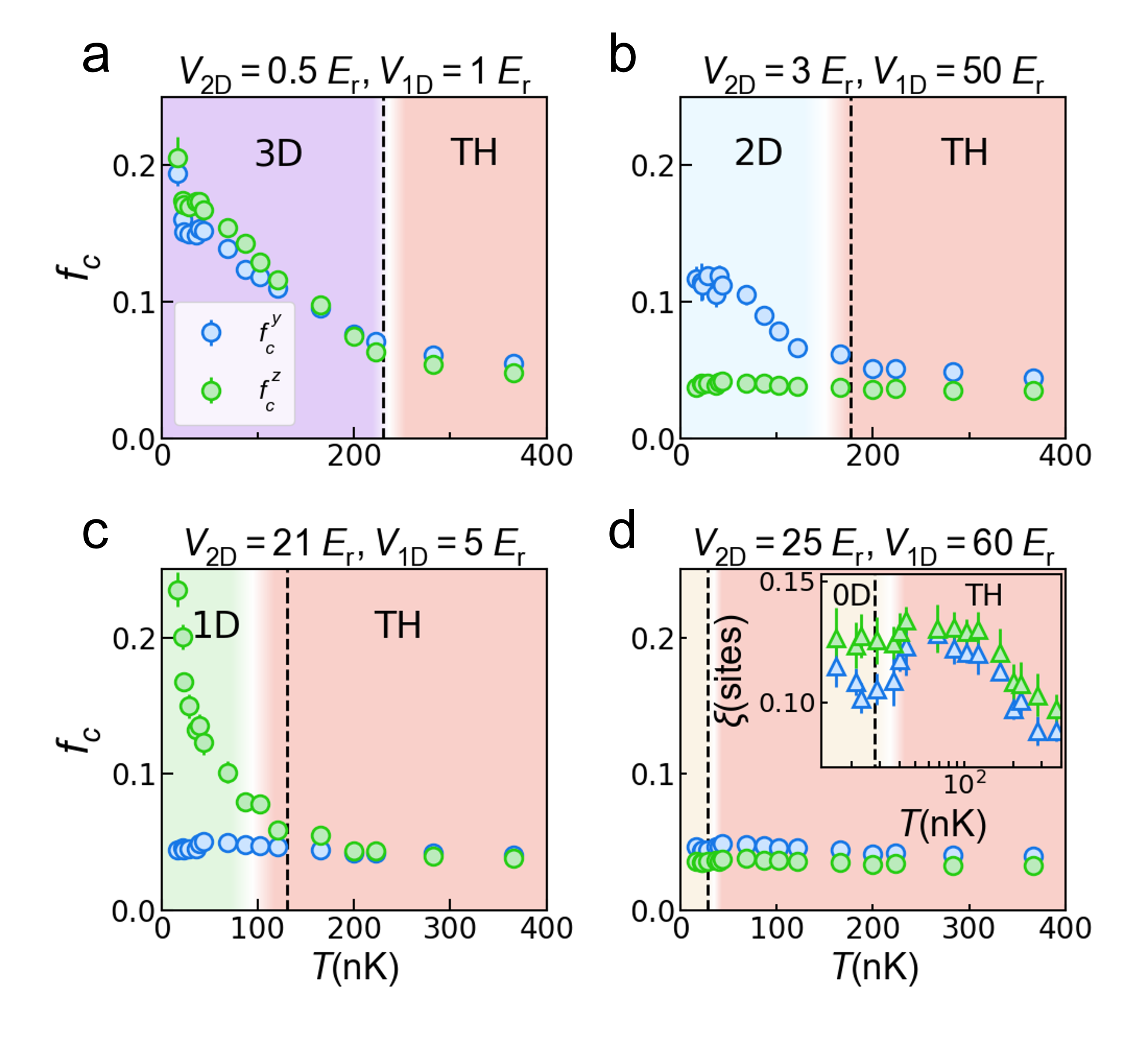}
    \caption{\textbf{The common quantum-to-thermal transition for different integer dimensionalites}
    The behavior of zero-momentum fraction $f_c$ along the y (blue) and z (green) directions as a function of temperature $T$. At the lowest temperature, the system is a quantum gas in the 3D (a, $V_{\rm 2D}=0.5 ~{E_{\rm r}}$, $V_{\rm 1D}=1 ~{E_{\rm r}}$), 2D (b, $V_{\rm 2D}=3.0 ~{E_{\rm r}}$, $V_{\rm 1D}=50.0 ~{E_{\rm r}}$), 1D (c, $V_{\rm 2D}=21.0 ~{E_{\rm r}}$, $V_{\rm 1D}=5.0 ~{E_{\rm r}}$) and 0D (d, $V_{\rm 2D}=25.0 ~{E_{\rm r}}$, $V_{\rm 1D}=60.0 ~{E_{\rm r}}$) regimes, respectively.  Error bars represent the standard deviation of five measurements. 
    The white region is the estimated transition temperature from the experimental data and the dashed lines are theoretical predictions.
    %The region indicated by white gradient color serve as guides for the crossover between different phases.  %and the statistical errorbar is smaller than the symbol size. 
    The inset of (d) shows the correlation length $\xi$ as a function of $T$. Experimental parameters: particle number $N = 2.0(3) \times 10 ^5$ and 3D s-wave scattering length $a_{3D} = 107(4) a_0$.}
    \label{fig4}
\end{figure}

In order to further study the properties of the quantum-to-thermal transition, we now choose six typical points (\uppercase\expandafter{\romannumeral1}-\uppercase\expandafter{\romannumeral6}) in Fig.~\ref{fig2}(a1), and scan $f^y_c$ and $f^z_c$ as a function of temperature $T$ while maintaining the particle numbers $N$ almost unchanged. The results are shown in Fig.~\ref{fig3} and Fig.~\ref{fig4}. Notably, here we use $f^y_c$ to study the coherence properties in the 2D $xy-$plane thanks to the rotational symmetry of the triangular lattices (see Supplementary Information). 

%Now, we turn to a more detailed study of the quantum-to-thermal transition across our parameter space. We choose six typical points in Fig.\ref{fig2}(a1) and scan quantum coherence properties by measuring the zero-momentum fraction along $y$ and $z$ directions (blue and green circles) as a function of temperature while maintaining the particle numbers almost unchanged, see Fig.\ref{fig3} and Fig.\ref{fig4}. Notably, here we use $f^y_c$ to study the coherence properties in the 2D $xy-$ plane thanks to the rotational symmetry of the triangluar lattices (see appendix). 

Although most of the cases fall into typical universality classes of quantum-to-thermal phase transition, there are some special points which show strikingly different behaviors, for instance points \uppercase\expandafter{\romannumeral1} and \uppercase\expandafter{\romannumeral2} in Fig.~\ref{fig2}(a1). Their finite temperature properties are shown in Fig.~\ref{fig3}. In Fig.~\ref{fig3}(a1), we consider the lattice depth as point  \uppercase\expandafter{\romannumeral1} and scan temperature. Interestingly, we find that $f^y_c$ and $f^z_c$ drop to a plateau at different values of temperature, namely $T_1=88 \pm 28 ~{\rm nK}$ and $T_2=142 \pm 14 ~{\rm nK}$ (white lines). This suggests, instead of a direct transition from 3D quantum to thermal phase, an intermediate 1D quantum regime emerges in between. It can be viewed as thermal fluctuation induced dimensional crossover, such that we name it "TFDC" type. 
Similar behavior has also been observed for another anisotropy case where $V_{\rm 1D}$ is large, see Fig.~\ref{fig3}(b1). The two $f_c$ drop to the plateau at two different temperatures, namely $T_1=119 \pm 19 ~\rm{nK}$ and $T_2=177 \pm 20 ~\rm{nK}$ (white lines). Similarly, it suggests a 3D-2D-TH process. 

The TFDC behavior will happen if the system has certain anisotropy \ie, lattice amplitudes along directions $j$ and $j'$ have a certain ratio. 
How should we understand it? As suggested by theoretical works~\cite{Cazalilla_2006,yao-crossoverD-2022, Pizzino2024}, the temperature competes with the effective hopping amplitude, renormalized by the interaction,  leading to the dimensional crossover. When it reaches the smaller hoping amplitude $t_j$, the particles are driven by thermal fluctuations along that direction. Effectively, we have one less direction where the system behaves coherently, see illustration in Fig.~\ref{fig3}(a2) and (b2). Therefore, as long as it happens below the thermal transition temperature, increasing temperature only eliminates the quantum coherence along $j$ direction but not the others like $j'$. This leads to a crossover from 3D quantum system to thermal regime via a low-D quantum phase. 

Notably, in order to probe these behaviors, one always needs to carefully pick up a point nearby the crossover line between the 3D and low-D regimes at low temperature, such that the temperature of dimensional crossover is much lower than the one to thermal phase. 
Although such mechanism was proposed in condensed matter systems, detecting it precisely is challenging given the difficulty to controlling the anisotropy accurately.
Thanks to the high tunability of parameters in our triangular lattice platform, we provide the first controlled test of this phenomenon.

To further confirm our demonstration, we perform two additional analyzes.
First, we estimate the 3D-1D crossover temperature via field theory~\cite{Pizzino2024}. By treating the system as coupled quantum chains, we can perform a mean-field (MF) decoupling estimate the crossover temperature~\cite{Cazalilla_2006,Pizzino2024}
\begin{equation}
     T_{{\textrm{\scriptsize 3-1D}}}= A_B t_{\perp}^{-\nu}.
    \label{T_3d-1d}
\end{equation}
with $\nu = \frac{2K}{4K-1}$ the scaling exponent and $K$ the Luttinger parameter encoding the effect of interactions. $A_B$ is the prefactor which depends on Luttinger parameters $K$, particle density $n$ and system size $L$. Using the experimental parameters, we find the temperature to be $T_{{\textrm{\scriptsize 3-1D}}}=108 ~\rm{nK}$ (purple dashed line), which fits with $T_1$ within errorbars in Fig.~\ref{fig3}(a1).

Another proof is the data of the one-body correlation function $G^{(1)}(r)=\int \langle \Psi^{\dagger} (r) \Psi (r') dr'$, which can be computed by the Fourier transform of the measured momentum distribution. It reflects clearly the correlation decay pattern along single directions. In Fig.~\ref{fig3}(a3), we show the decay of $G^{(1)}(y)$ and $G^{(1)}(z)$ at different temperatures. Clearly, above $T_1=88 \pm 28 ~\rm{nK}$, the correlation along $y$ direction drops extremely fast (faster than 1 site) and remains unchanged for higher temperatures. On the contrary, along the $z$ direction, the correlation drops faster while temperature increases, and only remains unchanged after $T_2 =142 \pm 14 ~\rm{nK}$. This further confirms the existence of the 3D-1D-TH transition at these two temperatures. In Fig.~\ref{fig3}(b3), similar behaviors are found for case \uppercase\expandafter{\romannumeral2}, where a signature of 3D-2D-TH transition is presented.

Now, we turn back to discuss the other common cases of the quantum-to-thermal transitions. In Fig.~\ref{fig2}(a2), cases like \uppercase\expandafter{\romannumeral1} and  \uppercase\expandafter{\romannumeral2} are minority. The majority should fall into four different universality class of phase transitions, namely the BEC transition (3D), BKT transition (2D), TLL transition (1D) and Mott melting (0D). Here, we pick up four points deeply in the regimes of 3D (\uppercase\expandafter{\romannumeral3}), 2D ( \uppercase\expandafter{\romannumeral4}), 1D (\uppercase\expandafter{\romannumeral5}) and 0D (\uppercase\expandafter{\romannumeral6}), and study the behavior of measured $f^y_c$ and $f^z_c$ as a function of temperature $T$, see Fig.~\ref{fig4}(a)-(d), correspondingly.

In Fig.~\ref{fig4}(a), the system starts from a 3D BEC at low temperature. By increasing temperature, both 
$f^y_c$ and $f^z_c$ drop together and reach a plateau at an identical value of $T_{\rm 3D}^{exp}=236 \pm 17$ nK. 
Here, the transition temperature for
a trapped 3D BEC writes~\cite{1995BEC}
\begin{equation}
    T_{\rm 3D}=T_{\rm BEC}= 0.94\frac{\hbar\overline{\omega}N^{1/3}}{k_{\rm B}},
    \label{T_bec}
\end{equation}
with $N$ the atom number, $\overline{\omega}=(\omega_x\omega_y\omega_Z)^{1/3}$ the average trapping frequency of the optical dipole trap, and $k_B$ the Boltzmann constant. It gives $T_{\rm BEC}=230~\rm{nK}$ (black dashed line) with our experimental parameters and fit nicely with our measured $T_{\rm 3D}^{exp}$ within $2.6 \%$.

For the 2D case, $f^z_c$ is always very small regardless of the value of temperature, while $f^y_c$ drops at $T_{\rm 2D}^{exp}=151 \pm 23$ nK and then remains constant, see Fig.~\ref{fig4}(b). This transition should be captured by the BKT transition class, whose transition temperature can be computed as~\cite{PhysRevLett.87.270402,PhysRevLett.100.140405}
\begin{equation}
    n_{\rm 2D}\lambda^2_{T_{\rm 2D}}=\ln{(\xi/4\pi)}+\ln{\ln{(1/n_{\rm 2D}a^2_{\rm 2D})}}
    \label{T_BKT}
\end{equation}
where $n_{\rm 2D}$ is the 2D density, $\lambda_{T} = \sqrt{h^2/2\pi mk_{B}T}$ is the thermal de Broglie wavelength, $a_{\rm 2D}$ is the 2D scattering length. The numerical coefficient $\xi=380$ is calculated in~\cite{PhysRevLett.87.270402}. Here, we find $T_{\rm BKT}= 176.3 ~\rm{nK}$ (black dashed line) which fits with the observed temperature $T_{\rm 2D}^{exp}$ within $15.5 \%$. Notably, for the case in Fig.~\ref{fig3}(b1), we can also estimate the 2D-TH transition temperature with Eq. (\ref{T_BKT}). It gives $T_{\rm 2D}=172.6 ~\rm{nK}$ (black dashed line) which fits nicely with the experimental observation within $2.5 \%$.

For the 1D case, while temperature increases, we find $f^y_c$ remains a small constant while $f^z_c$ decreases up to $T_{\rm 1D}^{exp}=95 \pm 23$ nK, see Fig.~\ref{fig4}(c).
This transition can be captured by the Tomonaga-Luttinger liquid theory~\cite{PhysRevB.82.060515,giamarchi_book_1d}. Specifically, its transition temperature writes
\begin{align}
    \xi (T=T_{\rm 1D}) =  \frac{3\hbar^2 \overline{n}_{\mathrm{1D}}}{2mk_{B}T_{\rm 1D}} \ll L
    \label{T_Lutt}
\end{align}  
with $\xi$ the correlation length, $n_{\rm 1D}$ the 1D density and $L$ the system size. 
In practice, we take $\xi(T=T_{\rm 1D})=L/10$ and it predicts $T_{\rm 1D}=130.9 ~\rm{nK}$ (black dashed line), which fits with our observation within $31.8 \%$. Similarly, we can apply this to the 1D-TH transition in Fig.~\ref{fig3}(a1). It gives $T_{\rm 1D}=155.5 ~\rm{nK}$  (black dashed line), which fits with our observation within $8.7 \%$ .

The finite temperature effect for a 0D quantum system corresponds to the melting effect discussed above. Both $f^y_c$ and $f^z_c$ are small at low temperature and remain almost unchanged with temperature, see Fig.~\ref{fig4}(d). To further capture this melting effect, we obtain the correlation function $G_1(x)$ from the measured $n(k)$, and by performing an exponential fit $G_1(x)\sim \rm{exp}(-|x|/\xi)$, we extract the correlation length $\xi$. The temperature dependence of $\xi$ is shown in the inset. At low temperature, it exhibits a plateau. Above the melting temperature $T_{\rm 0D}^{exp}=35 \pm 8 ~\rm{nK}$, it increases with temperature first, reaches a maximum value and then decreases. %Interestingly, this suggests an enhancement of coherence when temperature increases. 
We argue that this stems from the competition between the increasing mobility of particles induced by particle-hole pair excitations, and the increasing thermal fluctuations.
The scale of the melting temperature writes~\cite{gerbier2007}
\begin{equation}
    T_{\rm 0D} = T_{\rm melt}\sim \frac{ \Delta}{k_B},
    \label{T_melt}
\end{equation}
with $\Delta$ the Mott gap. Taking the parameters of our lattices, we find $\Delta=57 ~\rm{nK}$. In practice, we take $T_{\rm melt}= 0.4 \Delta/k_B = 28 ~\rm{nK}$ and find good agreement with experiment in both Fig.~\ref{fig2} and Fig.~\ref{fig4}(d). %\think{[HY: The 0D-TH transition needs to be investigated further.]}. 
%In fact, our measured zero-momentum fraction $f_c$ also exhibits such a non-monotonic behavior with a variation of $10\%$, which is too small to be visible.

%%%%%%%%%%%%%%%%%%%%%%%%%%%%%%%%%
%\section{Conclusion}

Summarizing, we probe the universal phase diagram of dimensional crossover for a quantum simulator at various temperatures, using interacting ultracold atoms in 3D anisotropic optical lattices. At various temperatures, we find quantum systems at different integer dimensionalities with a thermal phase existing in between. Furthermore, we study the quantum-to-thermal transition for systems with fixed anisotropy. Five categories of transitions are identified. Especially, we provide the first controlled test of the TFDC type, benefiting from the high tunability of both temperature and anisotropy in our triangular lattice setup. Our result provides important basis for quantum simulators with unconventional dimensionalities. Given such a platform, one can potentially carry out various further detailed tests, such as what is the mixed dimensional properties at various temperatures.
This kind of test also paves way to the understanding of their projective structures in condensed matter systems, especially the organic conductors and high-temperature superconductors.

\vspace{0.3cm}

%%%%%%%%%%%%%%%%%%%%%%%%%%%%%%%%%%%%%
\textbf{Acknowledgements} 

The authors thank Tianwei Zhou and Zekai Chen for their helpful discussions. This work is supported by the National Key Research and Development Program of China (Grants No. 2021YFA0718300 and No. 2021YFA1400900), National Natural Science Foundation of China (Grants No. 92365208). This work is also supported by the Swiss National Science Foundation under grant number 200020-219400.
\

\textbf{Author Contributions:} The work was conceived by J.T., H.Y., X.Z., T.G. and Z.Y. Experiments were performed by J.T., Z.Y and C.W. Data were analyzed by J.T. and J. L. Theoretical models and simulation were done by J.L., C.L., and L.P. Experiment preparations were done by Z.Y., C.W.,and H.S. X.Z., H.Y. and T.G. supervised this work. H.Y., J.T., Z.Y., T.G., C.L. wrote the manuscript with input from all authors. All authors discussed the results. 
\

\textbf{Data Availability:} The data shown in this manuscript is available via Zenodo~\cite{tian_2025_15308183}.

{\center \bf \large Methods \\}
\section{Control of temperature}
Our experiment starts from a Rubidium-87 Bose-Einstein Condensate (BEC) with a typical atomic number of $2.0(3) \times 10^5$ in the hyperfine state $|F=1, m_F=-1\rangle$, as shown in Fig.~\ref{fig1}. To further control the temperature of the produced BEC, we adjust the parameters of the evaporative cooling sequence as well as the initial state before this process. On the one hand, we prepare systems with different atom numbers before the evaporative cooling. This can be achieved by adjusting the magneto-optical trap (MOT) loading time prior to the evaporative cooling. On the other hand, we control the evaporative cooling sequence by varying both the decreasing rate and the final laser intensity. Combining these two processes properly, we can reach different temperatures while maintaining the same final atom number within a 15\% difference. In our measurement, the temperature of the BEC can be adjusted between 23 nK and 455 nK, as inferred from the TOF images using bimodal fitting~\cite{PhysRevLett.126.035301}.

%we vary both the decreasing rate during the cooling process and the final laser intensity. The corresponding trapping frequencies are tuned between $(\omega_x,\omega_y,\omega_z)/2\pi$ (27,84,80) and (60,135,121) Hz. To maintain a consistent atomic number across different temperatures, we adjust the magneto-optical trap (MOT) loading time before evaporative cooling.During the evaporative cooling, the atoms are evaporated out of the system with the decreasing of the trapping potential amplitudes, which reduces the temperature. Hence, in principle, when the final temperature decreases, the final atomic number should also decrease. Here, to maintain the same final atom number while varying the final temperature, we rather prepare systems with different atom numbers before the evaporative cooling. This can be done by changing the MOT loading time. With the proper control of the initial atom number before evaporative cooling, we can reach different final temperature while maintaining the same final atom number.
 
\section{The optical lattice potential}
Here, we clarify the details about the 1D and 2D lattice potentials. In the 2D $xy-$plane, our triangular lattice is formed by three traveling beams that intersect at an enclosing angle of $120^{\circ}$, with their linear polarization perpendicular to the 2D plane. The generated triangular lattice potential in the $xy-$plane is given by~\cite{Jin_2019}:
\begin{align}
    &V(x,y) =  -|E_1 + E_2 + E_3|^2 \nonumber \\
    &= -|\frac{\sqrt{V_{\rm{2D}}}}{2} e^{-i\vec{k}_1\cdot\vec{r}} + \frac{\sqrt{V_{\rm{2D}}}}{2} e^{-i\vec{k}_2\cdot\vec{r}} + \frac{\sqrt{V_{\rm{2D}}}}{2} e^{-i\vec{k}_3\cdot\vec{r}}|^2 \nonumber \\
    &=  -\frac{V_{\rm{2D}}}{4}(3 + 2\cos((\vec{k}_1 - \vec{k}_2)\cdot\vec{r}) \nonumber \\
    &\quad \ + 2\cos((\vec{k}_2 - \vec{k}_3)\cdot\vec{r}) + 2\cos((\vec{k}_3 - \vec{k}_1)\cdot\vec{r})) \nonumber \\ 
    &=  -\frac{V_{\rm{2D}}}{4}(3 + 2\cos(k_0\sqrt{3}x) + 4\cos(k_0\frac{\sqrt{3}}{2} x) \cos(k_0\frac{3}{2} y))
\end{align}
where $V_{\rm{2D}}$ is the lattice depth of 2D triangular lattice, $k_1$,$k_2$,$k_3$ are the wave vectors of three lattice beams. In our experiment, we always have $\vec{k}_1 = \frac{2\pi}{\lambda}(\frac{\sqrt{3}}{2},-\frac{1}{2}),\vec{k}_2 = \frac{2\pi}{\lambda}(-\frac{\sqrt{3}}{2},-\frac{1}{2}),\vec{k}_3 = \frac{2\pi}{\lambda}(0,1)$, and $|\vec{k}_1|=|\vec{k}_2|=|\vec{k}_3|=k_0$. Along the $z$-direction, we also load a 1D optical lattice formed by a $1064~{\rm nm}$ standing wave light. This potential can be expressed as $V(z) = V_{\rm{1D}} \cos^2{(k_0z)}$, where $V_{\rm{1D}}$ is lattice depth of 1D lattice, and $k_0 = \frac{2\pi}{\lambda}$ the wave vector with $\lambda = 1064 ~{\rm nm}$ the laser wavelength.

\section{The quantum Monte Carlo calculations} 
Using the quantum Monte Carlo method with worm algorithm~\cite{PhysRevLett.96.070601,PhysRevE.74.036701}, we simulate our experimental system based on the Bose-Hubbard model description at finite temperatures. Taking different temperature $T$, chemical potential $\mu$, on-site interaction $U$, and tunneling $t_{x,y,z}$ along three directions, we calculate the superfluid fraction along $i-$direction $f_s^i=\rho_s^i/\rho$ ($i=x,y,j$) by:
\begin{align}
f_{s,i} = \frac{m}{\hbar^2} \frac{\langle W_i^2 \rangle L_i^{2-d}}{\rho d \beta},
\end{align}
where $W_i$ is the winding number along $i$ direction, $L_i$ is the corresponded system size, $d$ is the total dimensionality of the simulation and $\beta = 1/k_{B} T$ is the inverse temperature.  For generating the phase diagrams in the main text (Fig.~\ref{fig2}(a2)-(d2)), we compute the superfluid fraction $f_s$ as a function of lattice depth $V$ and temperature $T$.
Typically, we perform $10^5$ iterations with $10^6$ warmup steps in advance, in order to make sure the Monte Carlo statistics is sufficient. The error bars of the QMC data originates
from the statistical fluctuations of the sampling.
Then, we determine the transition point as discussed in Fig.~\ref{fig1}(a3) and Fig.~\ref{fig4}. In practice, we define the criteria $f_s<0.1\%$~\cite{yao-crossoverD-2022}. 

\normalem
\bibliographystyle{revtex}
\bibliography{main}

\begin{thebibliography}{10}
	\providecommand*{\bibinfo}[2]{#2}
	\providecommand*{\eprint}[1]{#1}
	\providecommand*{\url}[1]{#1}
	\bibitem{wald2010general}
	\bibinfo{author}{R.~M. Wald}, \bibinfo{title}{\emph{General relativity}}
	(\bibinfo{publisher}{University of Chicago press}, \bibinfo{year}{2010}).
	\bibitem{akkermans2013}
	\bibinfo{author}{E.~Akkermans}, \bibinfo{title}{\emph{Statistical mechanics and
			quantum fields on fractals}}, \bibinfo{journal}{Contemp. Math.}
	\bibinfo{volume}{\textbf{601}}, \bibinfo{pages}{1} (\bibinfo{date}{2013}).
	\bibitem{jagannathan2021fibonacci}
	\bibinfo{author}{A.~Jagannathan}, \bibinfo{title}{\emph{The fibonacci
			quasicrystal: Case study of hidden dimensions and multifractality}},
	\bibinfo{journal}{Rev. Mod. Phys.} \bibinfo{volume}{\textbf{93}},
	\bibinfo{pages}{045001} (\bibinfo{date}{2021}).
	\bibitem{PATHRIA2022}
	\bibinfo{author}{R.~Pathria} and \bibinfo{author}{P.~D. Beale}, in
	\emph{Statistical Mechanics (Fourth Edition)} (\bibinfo{publisher}{Academic
		Press}, \bibinfo{year}{2022}).
	\bibitem{doi:10.1126/science.288.5465.468}
	\bibinfo{author}{J.~Orenstein} and \bibinfo{author}{A.~J. Millis},
	\bibinfo{title}{\emph{Advances in the physics of high-temperature
			superconductivity}}, \bibinfo{journal}{Science}
	\bibinfo{volume}{\textbf{288}}, \bibinfo{pages}{468} (\bibinfo{date}{2000}).
	\bibitem{doi:10.1021/cr030647c}
	\bibinfo{author}{T.~Giamarchi}, \bibinfo{title}{\emph{Theoretical framework for
			quasi-one dimensional systems}}, \bibinfo{journal}{Chem. Rev.}
	\bibinfo{volume}{\textbf{104}}, \bibinfo{pages}{5037} (\bibinfo{date}{2004}).
	\bibitem{CRPHYS_2024__25_G1_17_0}
	\bibinfo{author}{D.~Jerome} and \bibinfo{author}{C.~Bourbonnais},
	\bibinfo{title}{\emph{Quasi one-dimensional organic conductors: from
			{Fr\"ohlich} conductivity and {Peierls} insulating state to
			magnetically-mediated superconductivity, a retrospective}},
	\bibinfo{journal}{Comptes Rendus. Physique} \bibinfo{volume}{\textbf{25}},
	\bibinfo{pages}{17} (\bibinfo{date}{2024}).
	\bibitem{10.1093/oso/9780198562696.001.0001}
	\bibinfo{author}{M.~I. Eremets}, \bibinfo{title}{\emph{High Pressure
			Experimental Methods}} (\bibinfo{publisher}{Oxford University Press},
	\bibinfo{year}{1996}).
	\bibitem{PhysRevB.71.075104}
	\bibinfo{author}{M.~Dressel}, \bibinfo{author}{K.~Petukhov},
	\bibinfo{author}{B.~Salameh}, \bibinfo{author}{P.~Zornoza}, and
	\bibinfo{author}{T.~Giamarchi}, \bibinfo{title}{\emph{Scaling behavior of the
			longitudinal and transverse transport in quasi-one-dimensional organic
			conductors}}, \bibinfo{journal}{Phys. Rev. B} \bibinfo{volume}{\textbf{71}},
	\bibinfo{pages}{075104} (\bibinfo{date}{2005}).
	\bibitem{bloch-review-2008}
	\bibinfo{author}{I.~Bloch}, \bibinfo{author}{J.~Dalibard}, and
	\bibinfo{author}{W.~Zwerger}, \bibinfo{title}{\emph{Many-body physics with
			ultracold gases}}, \bibinfo{journal}{\Jrmp} \bibinfo{volume}{\textbf{80}},
	\bibinfo{pages}{885} (\bibinfo{date}{2008}).
	\bibitem{hadzibabic-2Dgas-2011}
	\bibinfo{author}{Z.~Hadzibabic} and \bibinfo{author}{J.~Dalibard},
	\bibinfo{title}{\emph{Two-dimensional {B}ose fluids: An atomic physics
			perspective}}, \bibinfo{journal}{Riv. Nuovo Cim.}
	\bibinfo{volume}{\textbf{34}}, \bibinfo{pages}{389} (\bibinfo{date}{2011}).
	\bibitem{cazalilla-1dreview-2011}
	\bibinfo{author}{M.~A. Cazalilla}, \bibinfo{author}{R.~Citro},
	\bibinfo{author}{T.~Giamarchi}, \bibinfo{author}{E.~Orignac}, and
	\bibinfo{author}{M.~Rigol}, \bibinfo{title}{\emph{One dimensional bosons:
			From condensed matter systems to ultracold gases}}, \bibinfo{journal}{\Jrmp}
	\bibinfo{volume}{\textbf{83}}, \bibinfo{pages}{1405} (\bibinfo{date}{2011}).
	\bibitem{paredes_tonks_experiment}
	\bibinfo{author}{B.~Paredes}, \bibinfo{author}{A.~Widera},
	\bibinfo{author}{V.~Murg}, \bibinfo{author}{O.~Mandel},
	\bibinfo{author}{S.~Folling}, \bibinfo{author}{I.~Cirac},
	\bibinfo{author}{{\relax GV}.~Shlyapnikov}, \bibinfo{author}{{\relax
			TW}.~Hansch}, and \bibinfo{author}{I.~Bloch},
	\bibinfo{title}{\emph{Tonks-{{Girardeau}} gas of ultracold atoms in an
			optical lattice}}, \bibinfo{journal}{Nature} \bibinfo{volume}{\textbf{429}},
	\bibinfo{pages}{277} (\bibinfo{date}{2004}).
	\bibitem{kinoshita2004}
	\bibinfo{author}{T.~Kinoshita}, \bibinfo{author}{T.~Wenger}, and
	\bibinfo{author}{D.~S. Weiss}, \bibinfo{title}{\emph{Observation of a
			one-dimensional {T}onks-{G}irardeau gas}}, \bibinfo{journal}{\Jscience}
	\bibinfo{volume}{\textbf{305}}, \bibinfo{pages}{1125} (\bibinfo{date}{2004}).
	\bibitem{giamarchi_book_1d}
	\bibinfo{author}{T.~Giamarchi}, \bibinfo{title}{\emph{Quantum physics in one
			dimension}}, \bibinfo{volume}{vol. 121 of \emph{International Series of
			Monographs on Physics}} (\bibinfo{publisher}{{Oxford University Press}},
	{Oxford}, \bibinfo{year}{2004}).
	\bibitem{goldman-topology-2013}
	\bibinfo{author}{N.~Goldman}, \bibinfo{author}{J.~Dalibard},
	\bibinfo{author}{A.~Dauphin}, \bibinfo{author}{F.~Gerbier},
	\bibinfo{author}{M.~Lewenstein}, \bibinfo{author}{P.~Zoller}, and
	\bibinfo{author}{I.~B. Spielman}, \bibinfo{title}{\emph{Direct imaging of
			topological edge states in cold-atom systems}}, \bibinfo{journal}{Proc. Natl.
		Acad. Sci. U.S.A.} \bibinfo{volume}{\textbf{110}}, \bibinfo{pages}{6736}
	(\bibinfo{date}{2013}).
	\bibitem{tarnowski-topology-2019}
	\bibinfo{author}{M.~Tarnowski}, \bibinfo{author}{F.~N. {\"U}nal},
	\bibinfo{author}{N.~Fl{\"a}schner}, \bibinfo{author}{B.~S. Rem},
	\bibinfo{author}{A.~Eckardt}, \bibinfo{author}{K.~Sengstock}, and
	\bibinfo{author}{C.~Weitenberg}, \bibinfo{title}{\emph{Measuring topology
			from dynamics by obtaining the {C}hern number from a linking number}},
	\bibinfo{journal}{Nat. Commun.} \bibinfo{volume}{\textbf{10}},
	\bibinfo{pages}{1728} (\bibinfo{date}{2019}).
	\bibitem{doi:10.1126/science.1207239}
	\bibinfo{author}{J.~Struck}, \bibinfo{author}{C.~{\"O}lschl{\"a}ger},
	\bibinfo{author}{R.~L. Targat}, \bibinfo{author}{P.~Soltan-Panahi},
	\bibinfo{author}{A.~Eckardt}, \bibinfo{author}{M.~Lewenstein},
	\bibinfo{author}{P.~Windpassinger}, and \bibinfo{author}{K.~Sengstock},
	\bibinfo{title}{\emph{Quantum simulation of frustrated classical magnetism in
			triangular optical lattices}}, \bibinfo{journal}{Science}
	\bibinfo{volume}{\textbf{333}}, \bibinfo{pages}{996} (\bibinfo{date}{2011}).
	\bibitem{hadzibabic2006}
	\bibinfo{author}{Z.~Hadzibabic}, \bibinfo{author}{P.~Kr{\"u}ger},
	\bibinfo{author}{M.~Cheneau}, \bibinfo{author}{B.~Battelier}, and
	\bibinfo{author}{J.~Dalibard},
	\bibinfo{title}{\emph{Berezinskii--kosterlitz--thouless crossover in a
			trapped atomic gas}}, \bibinfo{journal}{Nature}
	\bibinfo{volume}{\textbf{441}}, \bibinfo{pages}{1118} (\bibinfo{date}{2006}).
	\bibitem{ha-strong2Dboson-2013}
	\bibinfo{author}{L.-C. Ha}, \bibinfo{author}{C.-L. Hung},
	\bibinfo{author}{X.~Zhang}, \bibinfo{author}{U.~Eismann},
	\bibinfo{author}{S.-K. Tung}, and \bibinfo{author}{C.~Chin},
	\bibinfo{title}{\emph{Strongly interacting two-dimensional bose gases}},
	\bibinfo{journal}{\Jprl} \bibinfo{volume}{\textbf{110}},
	\bibinfo{pages}{145302} (\bibinfo{date}{2013}).
	\bibitem{dalibard-crossoverD-2023}
	\bibinfo{author}{G.~Chauveau}, \bibinfo{author}{C.~Maury},
	\bibinfo{author}{F.~Rabec}, \bibinfo{author}{C.~Heintze},
	\bibinfo{author}{G.~Brochier}, \bibinfo{author}{S.~Nascimbene},
	\bibinfo{author}{J.~Dalibard}, \bibinfo{author}{J.~Beugnon},
	\bibinfo{author}{S.~M. Roccuzzo}, and \bibinfo{author}{S.~Stringari},
	\bibinfo{title}{\emph{Superfluid fraction in an interacting spatially
			modulated {B}ose-{E}instein condensate}}, \bibinfo{journal}{Phys. Rev. Lett.}
	\bibinfo{volume}{\textbf{130}}, \bibinfo{pages}{226003}
	(\bibinfo{date}{2023}).
	\bibitem{spielman-crossover-2023}
	\bibinfo{author}{J.~Tao}, \bibinfo{author}{M.~Zhao}, and \bibinfo{author}{I.~B.
		Spielman}, \bibinfo{title}{\emph{Observation of anisotropic superfluid
			density in an artificial crystal}}, \bibinfo{journal}{Phys. Rev. Lett.}
	\bibinfo{volume}{\textbf{131}}, \bibinfo{pages}{163401}
	(\bibinfo{date}{2023}).
	\bibitem{Guo2024}
	\bibinfo{author}{Y.~Guo}, \bibinfo{author}{H.~Yao},
	\bibinfo{author}{S.~Ramanjanappa}, \bibinfo{author}{S.~Dhar},
	\bibinfo{author}{M.~Horvath}, \bibinfo{author}{L.~Pizzino},
	\bibinfo{author}{T.~Giamarchi}, \bibinfo{author}{M.~Landini}, and
	\bibinfo{author}{H.-C. N{\"a}gerl}, \bibinfo{title}{\emph{Observation of the
			2{D}--1{D} crossover in strongly interacting ultracold bosons}},
	\bibinfo{journal}{Nat. Phys.} \bibinfo{volume}{\textbf{20}},
	\bibinfo{pages}{934} (\bibinfo{date}{2024}).
	\bibitem{umesh-crossoverD-light-2024}
	\bibinfo{author}{K.~Karkihalli~Umesh}, \bibinfo{author}{J.~Schulz},
	\bibinfo{author}{J.~Schmitt}, \bibinfo{author}{M.~Weitz},
	\bibinfo{author}{G.~von Freymann}, and \bibinfo{author}{F.~Vewinger},
	\bibinfo{title}{\emph{Dimensional crossover in a quantum gas of light}},
	\bibinfo{journal}{Nat. Phys.} \bibinfo{volume}{\textbf{20}},
	\bibinfo{pages}{1810} (\bibinfo{date}{2024}).
	\bibitem{ho04_deconfined_bec}
	\bibinfo{author}{A.~Ho}, \bibinfo{author}{M.~Cazalilla}, and
	\bibinfo{author}{T.~Giamarchi}, \bibinfo{title}{\emph{Deconfinement in a
			{{2D}} optical lattice of coupled {{1D}} boson systems}},
	\bibinfo{journal}{Phys. Rev. Lett.} \bibinfo{volume}{\textbf{92}},
	\bibinfo{pages}{130405} (\bibinfo{date}{2004}).
	\bibitem{Cazalilla_2006}
	\bibinfo{author}{M.~A. Cazalilla}, \bibinfo{author}{A.~F. Ho}, and
	\bibinfo{author}{T.~Giamarchi}, \bibinfo{title}{\emph{Interacting bose gases
			in quasi-one-dimensional optical lattices}}, \bibinfo{journal}{New J. Phys.}
	\bibinfo{volume}{\textbf{8}}, \bibinfo{pages}{158} (\bibinfo{date}{2006}).
	\bibitem{PhysRevB.102.195145}
	\bibinfo{author}{G.~Bollmark}, \bibinfo{author}{N.~Laflorencie}, and
	\bibinfo{author}{A.~Kantian}, \bibinfo{title}{\emph{Dimensional crossover and
			phase transitions in coupled chains: Density matrix renormalization group
			results}}, \bibinfo{journal}{Phys. Rev. B} \bibinfo{volume}{\textbf{102}},
	\bibinfo{pages}{195145} (\bibinfo{date}{2020}).
	\bibitem{yao-crossoverD-2022}
	\bibinfo{author}{H.~Yao}, \bibinfo{author}{L.~Pizzino}, and
	\bibinfo{author}{T.~Giamarchi}, \bibinfo{title}{\emph{{Strongly-interacting
				bosons at 2D-1D dimensional crossover}}}, \bibinfo{journal}{SciPost Phys.}
	\bibinfo{volume}{\textbf{15}}, \bibinfo{pages}{050} (\bibinfo{date}{2023}).
	\bibitem{Pizzino2024}
	\bibinfo{author}{L.~Pizzino}, \bibinfo{author}{H.~Yao}, and
	\bibinfo{author}{T.~Giamarchi}, \bibinfo{title}{\emph{Finite size analysis
			for interacting bosons at the one-two dimensional crossover}},
	\bibinfo{journal}{Phys. Rev. Res.} \bibinfo{volume}{\textbf{7}},
	\bibinfo{pages}{013021} (\bibinfo{date}{2025}).
	\bibitem{Jin_2019}
	\bibinfo{author}{S.~Jin}, \bibinfo{author}{X.~Guo}, \bibinfo{author}{P.~Peng},
	\bibinfo{author}{X.~Chen}, \bibinfo{author}{X.~Li}, and
	\bibinfo{author}{X.~Zhou}, \bibinfo{title}{\emph{Finite temperature phase
			transition in a cross-dimensional triangular lattice}}, \bibinfo{journal}{New
		J. Phys.} \bibinfo{volume}{\textbf{21}}, \bibinfo{pages}{073015}
	(\bibinfo{date}{2019}).
	\bibitem{PhysRevLett.126.035301}
	\bibinfo{author}{S.~Jin}, \bibinfo{author}{W.~Zhang}, \bibinfo{author}{X.~Guo},
	\bibinfo{author}{X.~Chen}, \bibinfo{author}{X.~Zhou}, and
	\bibinfo{author}{X.~Li}, \bibinfo{title}{\emph{Evidence of potts-nematic
			superfluidity in a hexagonal $s{p}^{2}$ optical lattice}},
	\bibinfo{journal}{Phys. Rev. Lett.} \bibinfo{volume}{\textbf{126}},
	\bibinfo{pages}{035301} (\bibinfo{date}{2021}).
	\bibitem{Greiner2002}
	\bibinfo{author}{M.~Greiner}, \bibinfo{author}{O.~Mandel},
	\bibinfo{author}{T.~Esslinger}, \bibinfo{author}{T.~W. H{\"a}nsch}, and
	\bibinfo{author}{I.~Bloch}, \bibinfo{title}{\emph{Quantum phase transition
			from a superfluid to a mott insulator in a gas of ultracold atoms}},
	\bibinfo{journal}{Nature} \bibinfo{volume}{\textbf{415}}, \bibinfo{pages}{39}
	(\bibinfo{date}{2002}).
	\bibitem{PhysRevA.84.061606}
	\bibinfo{author}{T.~Plisson}, \bibinfo{author}{B.~Allard},
	\bibinfo{author}{M.~Holzmann}, \bibinfo{author}{G.~Salomon},
	\bibinfo{author}{A.~Aspect}, \bibinfo{author}{P.~Bouyer}, and
	\bibinfo{author}{T.~Bourdel}, \bibinfo{title}{\emph{Coherence properties of a
			two-dimensional trapped bose gas around the superfluid transition}},
	\bibinfo{journal}{Phys. Rev. A} \bibinfo{volume}{\textbf{84}},
	\bibinfo{pages}{061606} (\bibinfo{date}{2011}).
	\bibitem{Kato2008}
	\bibinfo{author}{Y.~Kato}, \bibinfo{author}{Q.~Zhou},
	\bibinfo{author}{N.~Kawashima}, and \bibinfo{author}{N.~Trivedi},
	\bibinfo{title}{\emph{Sharp peaks in the momentum distribution of bosons in
			optical lattices in the normal state}}, \bibinfo{journal}{Nat. Phys.}
	\bibinfo{volume}{\textbf{4}}, \bibinfo{pages}{617} (\bibinfo{date}{2008}).
	\bibitem{PhysRevA.71.063601}
	\bibinfo{author}{B.~DeMarco}, \bibinfo{author}{C.~Lannert},
	\bibinfo{author}{S.~Vishveshwara}, and \bibinfo{author}{T.-C. Wei},
	\bibinfo{title}{\emph{Structure and stability of mott-insulator shells of
			bosons trapped in an optical lattice}}, \bibinfo{journal}{Phys. Rev. A}
	\bibinfo{volume}{\textbf{71}}, \bibinfo{pages}{063601}
	(\bibinfo{date}{2005}).
	\bibitem{PhysRevLett.125.060401}
	\bibinfo{author}{H.~Yao}, \bibinfo{author}{T.~Giamarchi}, and
	\bibinfo{author}{L.~Sanchez-Palencia}, \bibinfo{title}{\emph{Lieb-liniger
			bosons in a shallow quasiperiodic potential: Bose glass phase and fractal
			mott lobes}}, \bibinfo{journal}{Phys. Rev. Lett.}
	\bibinfo{volume}{\textbf{125}}, \bibinfo{pages}{060401}
	(\bibinfo{date}{2020}).
	\bibitem{pitaevskii_becbook}
	\bibinfo{author}{L.~Pitaevskii} and \bibinfo{author}{S.~Stringari},
	\bibinfo{title}{\emph{Bose-Einstein Condensation}}
	(\bibinfo{publisher}{{Clarendon Press}}, {Oxford}, \bibinfo{year}{2003}).
	\bibitem{1995BEC}
	\bibinfo{author}{M.~H. Anderson}, \bibinfo{author}{J.~R. Ensher},
	\bibinfo{author}{M.~R. Matthews}, \bibinfo{author}{C.~E. Wieman}, and
	\bibinfo{author}{E.~A. Cornell}, \bibinfo{title}{\emph{Observation of
			{B}ose-{E}instein {C}ondensation in a dilute atomic vapor}},
	\bibinfo{journal}{Science} \bibinfo{volume}{\textbf{269}},
	\bibinfo{pages}{198} (\bibinfo{date}{1995}).
	\bibitem{PhysRevLett.87.270402}
	\bibinfo{author}{N.~Prokof'ev}, \bibinfo{author}{O.~Ruebenacker}, and
	\bibinfo{author}{B.~Svistunov}, \bibinfo{title}{\emph{Critical point of a
			weakly interacting two-dimensional bose gas}}, \bibinfo{journal}{Phys. Rev.
		Lett.} \bibinfo{volume}{\textbf{87}}, \bibinfo{pages}{270402}
	(\bibinfo{date}{2001}).
	\bibitem{PhysRevLett.100.140405}
	\bibinfo{author}{S.~Pilati}, \bibinfo{author}{S.~Giorgini}, and
	\bibinfo{author}{N.~Prokof'ev}, \bibinfo{title}{\emph{Critical temperature of
			interacting bose gases in two and three dimensions}}, \bibinfo{journal}{Phys.
		Rev. Lett.} \bibinfo{volume}{\textbf{100}}, \bibinfo{pages}{140405}
	(\bibinfo{date}{2008}).
	\bibitem{PhysRevB.82.060515}
	\bibinfo{author}{A.~Del~Maestro} and \bibinfo{author}{I.~Affleck},
	\bibinfo{title}{\emph{Interacting bosons in one dimension and the
			applicability of luttinger-liquid theory as revealed by path-integral quantum
			monte carlo calculations}}, \bibinfo{journal}{Phys. Rev. B}
	\bibinfo{volume}{\textbf{82}}, \bibinfo{pages}{060515}
	(\bibinfo{date}{2010}).
	\bibitem{gerbier2007}
	\bibinfo{author}{F.~Gerbier}, \bibinfo{title}{\emph{Boson mott insulators at
			finite temperatures}}, \bibinfo{journal}{\Jprl}
	\bibinfo{volume}{\textbf{99}}, \bibinfo{pages}{120405}
	(\bibinfo{date}{2007}).
	\bibitem{tian_2025_15308183}
	\bibinfo{title}{\emph{Data set is available from zenodo at doi:
			10.5281/zenodo.15308183.}}
	\bibitem{doi:10.1126/sciadv.adk6870}
	\bibinfo{author}{Y.~Guo}, \bibinfo{author}{H.~Yao}, \bibinfo{author}{S.~Dhar},
	\bibinfo{author}{L.~Pizzino}, \bibinfo{author}{M.~Horvath},
	\bibinfo{author}{T.~Giamarchi}, \bibinfo{author}{M.~Landini}, and
	\bibinfo{author}{H.-C. N{\"a}gerl}, \bibinfo{title}{\emph{Anomalous cooling
			of bosons by dimensional reduction}}, \bibinfo{journal}{Sci. Adv.}
	\bibinfo{volume}{\textbf{10}}, \bibinfo{pages}{eadk6870}
	(\bibinfo{date}{2024}).
	\bibitem{PhysRevA.91.043617}
	\bibinfo{author}{N.~Fabbri}, \bibinfo{author}{M.~Panfil},
	\bibinfo{author}{D.~Cl\'ement}, \bibinfo{author}{L.~Fallani},
	\bibinfo{author}{M.~Inguscio}, \bibinfo{author}{C.~Fort}, and
	\bibinfo{author}{J.-S. Caux}, \bibinfo{title}{\emph{Dynamical structure
			factor of one-dimensional bose gases: Experimental signatures of
			beyond-luttinger-liquid physics}}, \bibinfo{journal}{Phys. Rev. A}
	\bibinfo{volume}{\textbf{91}}, \bibinfo{pages}{043617}
	(\bibinfo{date}{2015}).
	\bibitem{PhysRevLett.115.085301}
	\bibinfo{author}{F.~Meinert}, \bibinfo{author}{M.~Panfil},
	\bibinfo{author}{M.~J. Mark}, \bibinfo{author}{K.~Lauber},
	\bibinfo{author}{J.-S. Caux}, and \bibinfo{author}{H.-C. N\"agerl},
	\bibinfo{title}{\emph{Probing the excitations of a lieb-liniger gas from weak
			to strong coupling}}, \bibinfo{journal}{Phys. Rev. Lett.}
	\bibinfo{volume}{\textbf{115}}, \bibinfo{pages}{085301}
	(\bibinfo{date}{2015}).
	\bibitem{PhysRevA.107.L061302}
	\bibinfo{author}{K.-Y. Li}, \bibinfo{author}{Y.~Zhang},
	\bibinfo{author}{K.~Yang}, \bibinfo{author}{K.-Y. Lin},
	\bibinfo{author}{S.~Gopalakrishnan}, \bibinfo{author}{M.~Rigol}, and
	\bibinfo{author}{B.~L. Lev}, \bibinfo{title}{\emph{Rapidity and momentum
			distributions of one-dimensional dipolar quantum gases}},
	\bibinfo{journal}{Phys. Rev. A} \bibinfo{volume}{\textbf{107}},
	\bibinfo{pages}{L061302} (\bibinfo{date}{2023}).
	\bibitem{PhysRevLett.96.070601}
	\bibinfo{author}{M.~Boninsegni}, \bibinfo{author}{N.~Prokof'ev}, and
	\bibinfo{author}{B.~Svistunov}, \bibinfo{title}{\emph{Worm algorithm for
			continuous-space path integral monte carlo simulations}},
	\bibinfo{journal}{Phys. Rev. Lett.} \bibinfo{volume}{\textbf{96}},
	\bibinfo{pages}{070601} (\bibinfo{date}{2006}).
	\bibitem{PhysRevE.74.036701}
	\bibinfo{author}{M.~Boninsegni}, \bibinfo{author}{N.~V. Prokof'ev}, and
	\bibinfo{author}{B.~V. Svistunov}, \bibinfo{title}{\emph{Worm algorithm and
			diagrammatic monte carlo: A new approach to continuous-space path integral
			monte carlo simulations}}, \bibinfo{journal}{Phys. Rev. E}
	\bibinfo{volume}{\textbf{74}}, \bibinfo{pages}{036701}
	(\bibinfo{date}{2006}).
	
\end{thebibliography}

%%%%%%%%%%%%%%%%%%%%%%%%%%%%%%%%%%%%%

 \renewcommand{\theequation}{S\arabic{equation}}
 \setcounter{equation}{0}
 \renewcommand{\thefigure}{S\arabic{figure}}
 \setcounter{figure}{0}
 \renewcommand{\thesection}{S\arabic{section}}
 \setcounter{section}{0}
 \onecolumngrid  
     
 %\pagebreak
 
 \newpage

 {\center \bf \large Supplemental Material for \\}
 {\center \bf \large Probing universal phase diagram of dimensional crossover with an atomic quantum simulator\\ \vspace*{1.cm}
 }

In this supplemental material, we provide details about the comparison between different lattice geometries, experimental details, determination of the critical potentials and the properties in the 2D planes.

\section{Comparison between different lattice geometries}

\begin{figure}[ht]
    \centering
    \includegraphics[width=0.4\columnwidth]{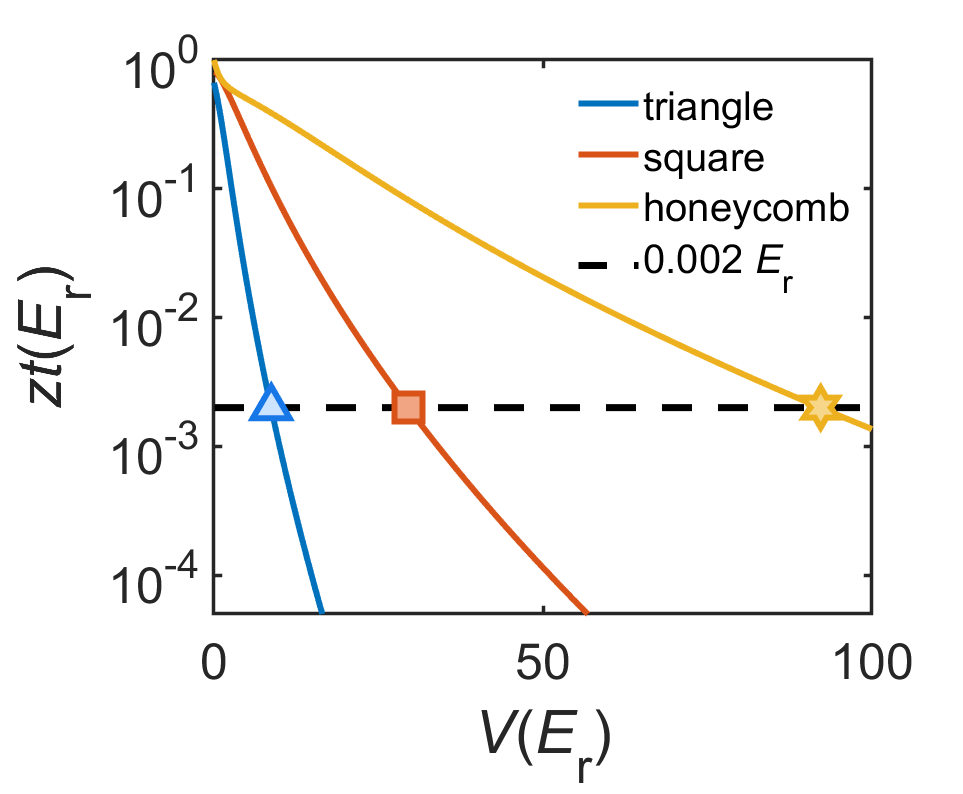}
    \caption{ \textbf{Plot of $z t-V$ for different lattice geometries.} The product of tunneling coefficient $t$ and coordination number $z$ as a function of the lattice depth $V$ is shown for the two-dimensional triangular (blue), square (red), and honeycomb lattices (yellow). The dashed line on the graph represents $zt =0.002 ~{E_{\rm r}}$. Its intersection with the $V-zt$ curve for the triangular lattice occurs at 8.5$~{E_{\rm r}}$, indicating the lattice depth at which the system undergoes a 3D-1D dimensional crossover at low temperatures. Based on this criterion, the critical lattice depths for the square and honeycomb lattices are found to be $29.5~{E_{\rm r}}$ and $92.2~{E_{\rm r}}$, respectively.
    }
\label{figs1} 
\end{figure}

In this section, we discuss the effect of different lattice geometries on the phase diagrams and verify the advantages of using a triangular lattice. As we argued in the main text, when we use different lattice structures with different nearest neighbor values $z$, the physics remains unchanged. For instance, the critical tunneling value of the 3D-1D crossover writes~\cite{Cazalilla_2006}
\begin{equation}
    t_c = \frac{1}{z} \alpha(\rho,v_s,m)   T^{2-1/2K},
    \label{zct}
\end{equation}
where pre-factor $\alpha(\rho_0,v_s,m)$ is the function of the sound velocity $v_s$, 1D density $\rho_0$, and atomic mass $m$. The temperature dependence of $t_c$ is a power law whose exponent is determined by the Luttinger parameter $K$. The product of $K$ and $v_s$ is fixed by Galilean invariance, as $v_sK=\hbar\pi\rho_0/m$~\cite{Cazalilla_2006}.

\iffalse
Notably, the transition points varies qualitatively and indeed when we carry out the experimental measurement of the phase diagram in Fig.~\ref{fig2} of the main text, we observe a significant difference. Since different lattice structures have different bandwidths of the ground band ($S$ band), if the $S$ band becomes flatter, the tunneling amplitude of the system generally decreases under tight-binding conditions. Given the same peak-to-peak potential energy of a lattice, the order of bandwidth $W$ for triangular, square and honeycomb lattice is $W_{\rm tri}<W_{\rm sq}<W_{\rm hon}$. Consequently, for a given potential amplitude $V$, the tunneling amplitude for these three lattice types in the 2D direction follows the order $t_{\rm tri}<t_{\rm sq}<t_{\rm hon}$.
\fi

For the triangular lattice, we observe a clear 3D to 1D crossover at 23 nK when $V_{\rm 2D}$ is approximately $8.5\ E_{\rm{r}}$, as shown in Fig.~\textcolor{red}{2} of the main text. According to Eq.~\ref{zct}, when the other parameters are the same, the 3D to 1D crossover occurs at the same $zt$ for different lattice geometries. For triangular lattices $(z=6)$ that we studied, we find $z t=0.002\ E_{\rm{r}}$ for $V_{\rm tri}=8.5\ E_{\rm{r}}$. In contrast, for square lattices ($z=4$) and honeycomb lattices ($z=3$), $zt=0.002\ E_r$ leads to $V_{\rm sq}=29.5\ E_{\rm{r}}$ and $V_{\rm hon}=92.2\ E_{\rm{r}}$, correspondingly, as shown in Fig.~\ref{figs1}. This suggests that much higher laser power is required to detect the full phase diagram of dimensions in square and honeycomb lattices. On the one hand, this may exceed the actual laser power limit of the experiment. On the other hand, such high laser power may heat transmissive optical components, causing laser intensity feedback fluctuations and beam waist displacement, ultimately leading to instability in lattice depths.

\section{Additional experimental details}

After preparing the BEC, it is loaded into the optical lattice, which divides the BEC into either 2D layers or 1D tubes, allowing us to investigate different dimensionalities in the system. During the loading process, the BEC exhibits a 3D density distribution $n(x,y,z)$ that follows the Thomas-Fermi (TF) profile and is normalized such that $\int \mathrm{d}x \, \mathrm{d}y \, \mathrm{d}z \, n(x,y,z) = N$, where $N$ is the total particle number. By projecting this 3D distribution onto the discretized lattice configuration $n(i,j,k)$, we can calculate the atom number in layers along the lattice sites for the constrained lattice $N_k$ (2D case) or in tubes allocated at lattice sites in the triangular lattice $N_{i,j}$ (1D case). The weighted atom numbers for the effective 2D layer or 1D tube are evaluated as:~\cite{Guo2024,doi:10.1126/sciadv.adk6870,PhysRevA.91.043617,PhysRevLett.115.085301,PhysRevA.107.L061302} 
\begin{align}
\overline{N}_{\rm 2D} = \frac{\sum_{k} N_k^2}{\sum_{k} N_k} \quad \text{and} \quad \overline{N}_{\rm 1D} = \frac{\sum_{i,j} N_{i,j}^2}{\sum_{i,j} N_{i,j}}.
\end{align}
In practice, we find that the weighted atom numbers for the effective 2D and 1D systems in our experiment are $\overline{N}_{\rm 2D} = 13,000$ and $\overline{N}_{\rm 1D} = 620$, respectively.

Moreover, for distinguishing the 0D and thermal phases, we need to measure the correlation length $\xi$. 
By applying Fourier transform to the momentum distribution $n(k)$ obtained from the TOF images, we obtain the integrated correlation function $G^{(1)}$ which can be written:
\begin{equation}
    G^{(1)} (x) = \int\frac{dx'}{L}\langle\Psi(x'+x)\Psi(x')\rangle
\end{equation}
For 1D atomic gases under the typical experimental condition, \ie at nano-Kelvin temperature scale and with the presence of a harmonic trap, the $G^{(1)}$ is usually dominated by an exponential decay~\cite{Guo2024}. This is also the case for all of our experimental measurements. 
Therefore, by fitting $G^{(1)}(x)\sim e^{-x/\xi}$, we obtain the correlation length at various temperatures and lattice depths, as shown in Fig.~4(d) of the main text. Such a curve allows us to distinguish the 0D phase from the thermal phase.

\begin{figure}[h!]
    \centering
    \includegraphics[width=0.6\columnwidth]{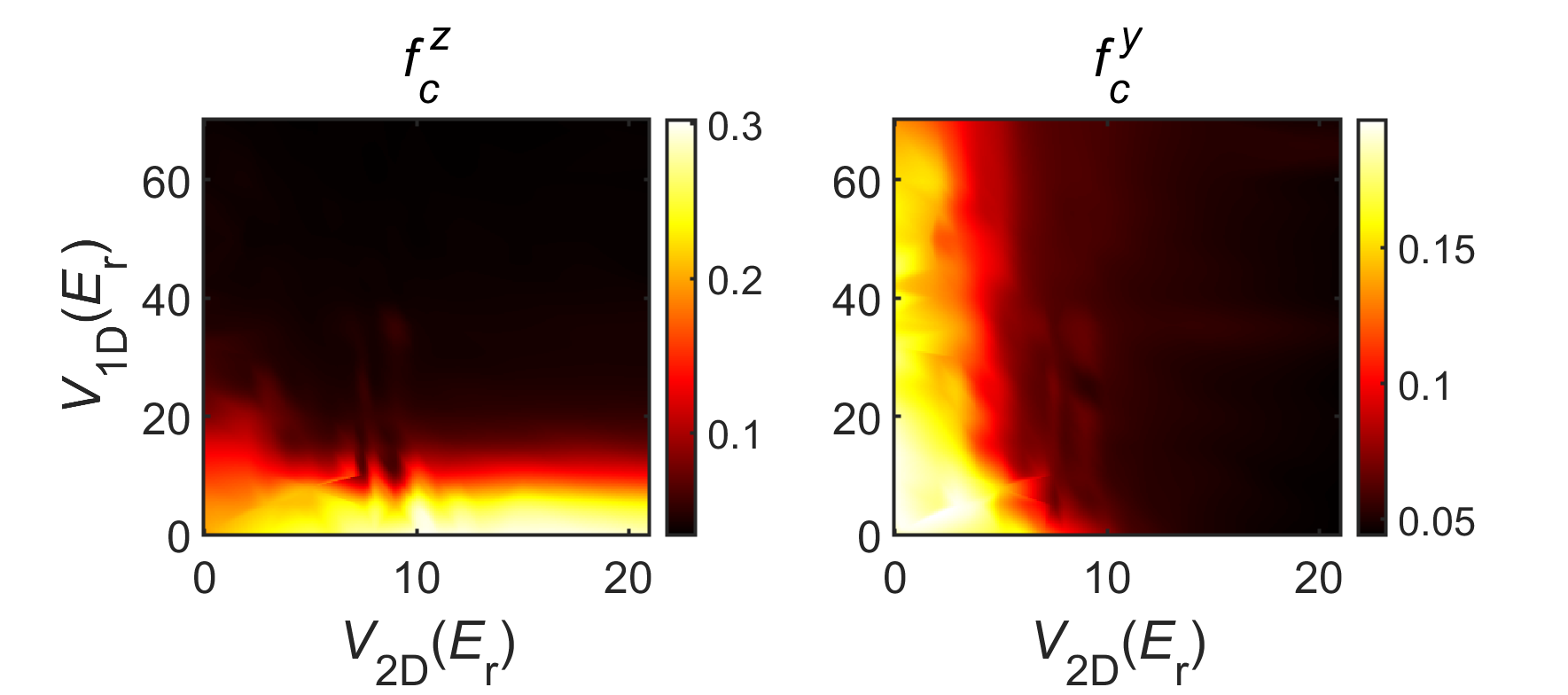}
    \caption{ \textbf{Zero-momentum fraction $f_c$ versus lattice depths.} The zero-momentum fraction in the z-direction $f_{c}^z$ (left panel) and y-direction $f_{c}^y$ (right panel) is displayed as a function of the lattice depths $V_{\rm 1D}$ and $V_{\rm 2D}$. The figure combines experimental data with interpolated values to provide a smooth presentation of the underlying trend. Experimental parameters: Temperature $T=16$ nK, $N = 2.0(3) \times 10 ^5$, 3D s-wave scattering length $a_{\rm 3D} = 107(4) a_0$. and the trap frequencies $(\omega_x,\omega_y,\omega_z)/2\pi=$ (27,84,80) Hz. 
 }
\label{figs2} 
\end{figure}

\section{Determination of the critical potentials}

To calculate the critical points for the dimensional crossover, we use the zero-momentum fraction $f^i_c$ along the $i$-direction as the criterion. According to Refs.~\cite{Guo2024, Pizzino2024}, this quantity is as effective as the superfluid fraction for determining the crossover point.

From the TOF images in our experiment, we can extract the zero-momentum fractions in the $y$- and $z$-directions, denoted by $f_{c}^y$ and $f_{c}^z$, which reflect the coherence in the 2D triangular lattice plane and the 1D constrained lattice direction, respectively. In Fig.~\ref{figs2}, we show typical examples of their behavior as functions of the lattice depths $V_{\rm 1D}$ and $V_{\rm 2D}$. Clearly, the zero-momentum fraction $f_{c}^j$ decays as the lattice depth along the corresponding direction $i$ increases. In each subfigure, we observe two distinct regions. At low lattice depths along the $j$-direction, $f_c^j$ remains high, but as $V_j$ increases, $f_c^j$ rapidly decays and then converges to a constant. Theoretically, the superfluid fraction should exhibit a similar qualitative behavior.

%We show the zero-momentum fraction in z direction $f_{c}^z$ (left pannel) and y direction $f_{c}^y$ (right pannel), as a function of the 1D and 2D lattice depths, $V_{1D}$ and $V_{2D}$. The colorbar indicates the zero-momentum fraction $f_c$.

\begin{figure}[h!]
    \includegraphics[width=0.6\columnwidth]{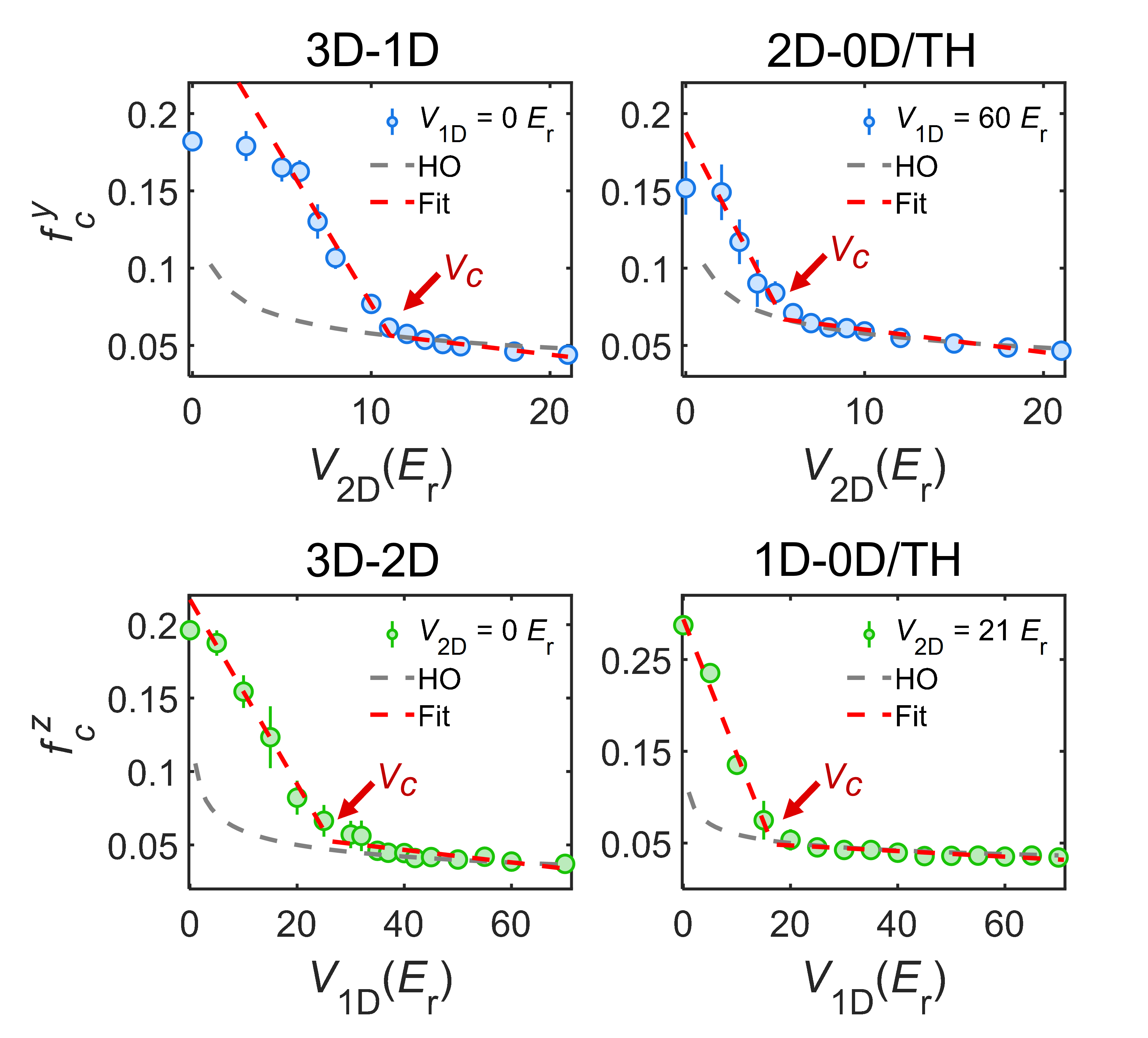}
    \caption{ \textbf{Typical $f_c-V$ curves across dimensional crossovers at low temperatures.} We present four typical dimensional crossover measurements by studying $f^y_c$ and $f^z_c$ as a function of both $V_{\rm 1D}$ and $V_{\rm 2D}$. The four plots correspond to: (1). 3D-1D crossover with $V_{\rm 1D}=0\ E_{\rm{r}}$; (2) 2D-0D/TH crossover with $V_{\rm 1D}=60\ E_{\rm{r}}$; (3) 3D-2D crossover with $V_{\rm 2D}=0\ E_{\rm{r}}$; (4) 1D-0D/TH crossover with $V_{\rm 2D} =21\ E_{\rm{r}}$. The red dashed line represents piecewise fitting, and the inflection point of the line marks the critical point of crossover. The gray dashed line represents the $f^y_c$ or $f^z_{c}$ calculated using the harmonic oscillator (HO) approximation for a single lattice site, where the deviation occurs from the data indicating the critical point. The error bars represent the standard deviation from five measurements. Experimental parameters: Temperature $T=16\ \rm nK$, $N = 2.0(3) \times 10 ^5$, 3D s-wave scattering length $a_{\rm 3D} = 107(4) a_0$, and the trapping frequencies $(\omega_x,\omega_y,\omega_z)/2\pi=$ (27,84,80) Hz.}
    \label{figs2_2} 
\end{figure}

Next, we focus on specific cuts of Fig.~\ref{figs2}, as illustrated by the four representative cases in Fig.~\ref{figs2_2}. At low lattice depths, increasing the depth exponentially reduces the tunneling coefficient between lattice sites, causing $f_c$ to decay rapidly. When the lattice depth exceeds the dimensional crossover point, the lattice sites become incoherently coupled, and the system can be treated as a collection of separate harmonic oscillators along the transverse direction. Further increasing the lattice depth only increases the trapping frequency, which in turn broadens the momentum distribution of the ground state. As shown in Fig.~\ref{figs2_2}, we apply the piecewise fitting to the $f_c$ curves to determine the critical lattice depth $V_c$.

Additionally, we calculate the zero-momentum fraction for an equivalent quantum harmonic oscillator, denoted as $f^{i}_{\rm harm}$, represented by the grey dashed lines~\cite{Guo2024}
\begin{align}
f^{i}_{\rm harm} = \frac{\int_{-\Delta k_i}^{-\Delta k_i}n_{\rm HO}(k_i)dk_i}{\int_{-\infty}^{+\infty}n_{\rm HO}(k_i)dk_i},\quad n_{\rm HO}(k_i) = (\frac{1}{\pi\hbar m \omega_i})^{1/2} \exp(-\frac{\hbar k^2}{m \omega_i}) 
\end{align} where $m$ and $\omega_i/2 \pi$ are the atomic mass and the transverse trapping frequency, respectively.
At low temperatures, the point where $f^{i}_c$ and $f_{\rm harm}$ begin to coincide corresponds to the critical point $V_c$, at which the superfluid fraction disappears~\cite{Guo2024}. For all subfigures in Fig.~\ref{figs2_2}, we observe that $f_{\rm harm}$ fits well with the experimental data when $V > V_c$, while the two curves separate when $V < V_c$. However, at higher temperatures, due to thermal excitation, atoms occupy excited states in addition to the ground state of the harmonic trap, causing $f^{i}_c$ to be lower than $f_{\rm harm}$ even at high lattice depths. Despite this, the two distinct decay regimes of $f_c$ remain apparent, allowing us to determine $V_c$ based on a piecewise fit.

\begin{figure*}[h!]
    \centering    \includegraphics[width=0.4\columnwidth]{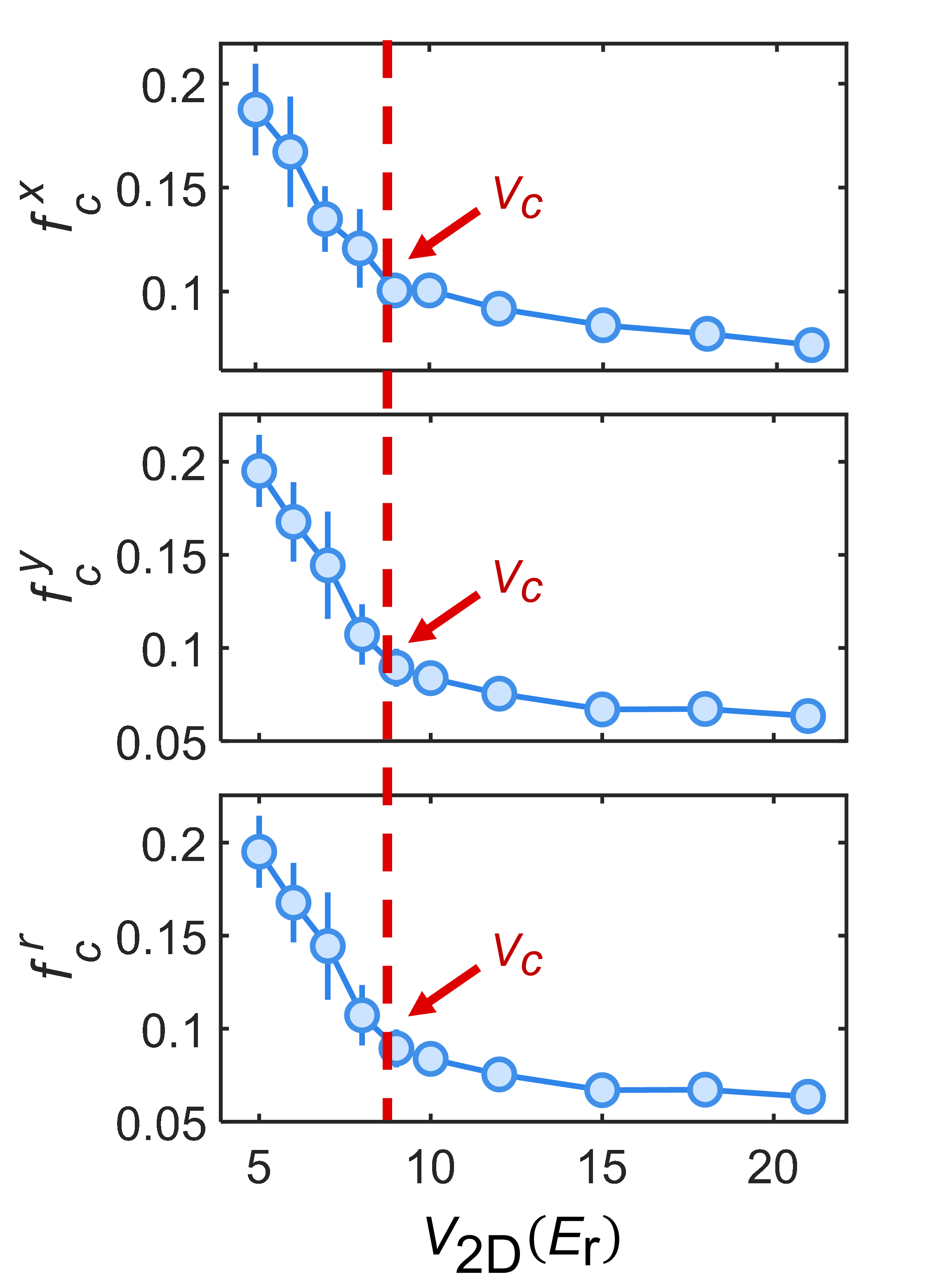}
    \caption{\textbf{$f_c-V_{\rm 2D}$ curves along $x$, $y$ directions and angular averaged.}
    %(a) TOF image in XY plane. Typical TOF images of 3D to 1D crossover in the xy plane. We fix $V_{1D}$ at $5 E_r$ and adjust $V_{2D}$ from $0 E_r$ to $21 E_r$. During the increase of $V_{2D}$, the momentum peak in the xy plane gradually becomes dispersed. 
    The blue circles show the zero-momentum fraction along x direction $f_c^x$, y direction $f_c^y$, and the angular averaged radius $f_c^r$. The red dashed lines represent the common critical point $V_c$ obtained from these three curves. The error bars represent the standard deviation of five measurements. Experimental parameters: Temperature $T = 16$ nK, $N = 2.0(3) \times 10^5$, 3D s-wave scattering length $a_{3D} = 107(4) a_0$.}
\label{figs4} 
\end{figure*}

\section{Properties in the 2D plane}

In this section, we verify why we can use the information along $y$ directions to represent the physics in the 2D $x-y$ plane. Thanks to the $C_6$ rotational symmetry of the triangular lattice, we expect $f_c$ along x and y directions to exhibit the same property. 
To prove this, we take TOF images in the $x-y$ plane for typical parameter values. One example is shown in Fig.~\ref{figs4}, where we take the 1D lattice depth $V_{\rm 1D}=5 ~E_{\rm{r}}$ and temperature $T=16 ~\rm{nk}$. We modify the 2D lattice depth to achieve a dimensional crossover from 3D to 1D. When $V_{\rm 2D}$ increases, the central atomic cluster gradually becomes dispersed showing no anisotropy, which suggests the superfluid along x and y directions disappear at the same time. 

%After time-of-flight(TOF) process and absorption imaging, we obtain momentum distribution in yz-plane $n(k_y,k_z) = \int_{-\infty}^{+\infty} n(k_x,k_y,k_z) dk_x$. z direction can reflect superfluid of 1D lattice, x and y direction can reflect superfluid of 2D lattice. But the momentum distribution in the x direction has been integrated during the imaging process. While the other direction of 2D triangular lattice is y direction, where simultaneously possessing or losing superfluidity with x direction because of the $C_6$ rotational symmetry. Thus, $f_c$ in x and y direction exhibit the same property. 

%\begin{figure*}[ht]
%    \centering
%    \includegraphics[width=0.4\columnwidth]{fig/fig_s4_2_7.png}
%   \caption{\textbf{$f_c-V_{2D}$ curves along $x$, $y$ directions and angular averaged} Zero-momentum fraction along $x$ direction $f_c^x$, $y$ direction $f_c^y$, and the angular averaged radius $f_c^r$. The error bars represent the standard deviation of five measurements. The red dashed line represents the common critical point $V_c$ obtained from these three curves. Experiment\todo{al} parameters: Temperature $T=16$ nK, $N \approx 2 \times 10^5$, 3D s-wave scattering length $a_{3D} = 107(4) a_0$, and the trap frequencies $(\omega_x,\omega_y,\omega_z)/2\pi=$ (27,84,80) Hz.}
%\label{figs4_2} 
%\end{figure*}

Furthermore, we use a set of experimental data to illustrate our statement quantitatively, as shown in Fig.~\ref{figs4}. Apparently, the $f_c$ line shapes in both $x$ and $y$ directions are consistent, with similar magnitudes, and the phase transition points are essentially identical. We also perform angular averaging of the TOF image to obtain the momentum distribution as a function of the scalar momentum $|\mathbf{k}|$, which allows us to calculate the angularly averaged zero-momentum fraction $f_c^r$. In Fig.~\ref{figs4} , we find that the behavior of $f_c^r$ is consistent with that of $f_c^x$ and $f_c^y$, in terms of the line shapes, magnitudes, and critical points. This proves our statement and assures that we can determine the phase transition points solely by measuring $f_c^y$ in the $y$-$z$ plane.

\end{document}